  \pdfoutput=1
  \documentclass[graphicx,amssymb,amsmath,enumerate,11pt]{report}
  \usepackage{fancyhdr}
  \fancyheadoffset[]{0.1cm}

   \usepackage{epsfig}
  \usepackage{url}
  \newcommand\mchapter[2]{\chapter*{#1}
  \vskip -0.5cm \noindent {\it \LARGE #2}
  \addcontentsline{toc}{chapter}{#1\\{\normalsize\it #2}}}

\newcommand{\be}{\begin{equation}}
\newcommand{\ee}{\end{equation}}
\newcommand{\bea}{\begin{eqnarray}}
\newcommand{\eea}{\end{eqnarray}}

\def\s12{\sin\theta_{12}}
\def\s23{\sin\theta_{23}}
\def\s13{\sin\theta_{13}}
\def\ts12{\theta_{12}}
\def\ta23{\theta_{23}}
\def\t13{\theta_{13}}

\begin{document}      

 \rhead{\bfseries Neutrino Yukawa textures within type-I see-saw}

 \mchapter{Neutrino Yukawa textures within type-I see-saw}
 {Authors:\ Biswajit Adhikary$^{a}$
 and Probir Roy$^{b}$}
 \label{ch-07:neutrinoYukawa}

\vspace{0.5cm}

\begin{center}
$^a$ {\it Department of Physics, Gurudas College,
Narkeldanga, Kolkata-700054, India\\
email: biswajitadhikary@gmail.com
} \\ [6pt]
$^b$ {\it Saha Institute of Nuclear Physics, 1/AF Bidhan
        Nagar, Kolkata 700064, India\\
email: probirrana@gmail.com}
\end{center}

\vspace{1cm}

\begin{center}
{\bf Abstract}
\end{center}
The arbitrariness of Yukawa couplings can be reduced by the imposition of some flavor symmetries and/or by the 
realization of 
texture zeros. We review neutrino Yukawa textures with zeros within the framework of the 
 type-I seesaw with three heavy right chiral neutrinos 
and in the basis where 
the latter and the charged leptons are mass diagonal. An assumed  non-vanishing mass of every ultralight neutrino
and the observed non-decoupling of any neutrino generation allow a maximum of four zeros  
in the Yukawa coupling matrix $Y_\nu$ in family space. There are seventy two such textures. We show that the requirement of an 
exact $\mu\tau$ symmetry, 
coupled with the observational constraints,
reduces these seventy two allowed textures to {\it only four} 
corresponding to {\it just two} different forms of the light neutrino mass matrix $M_{\nu A}/M_{\nu B}$, 
resulting in an inverted/normal mass ordering.
The effect of each of these on measurable quantities can be described, 
apart from an overall factor of the neutrino mass scale, 
in terms of two real parameters and a phase angle all of which are within very 
constrained ranges. The masses and Majorana phases of ultralight neutrinos
 are predicted within definite ranges with $3\sigma$ laboratory and cosmological observational inputs. 
The rate for $0\nu\beta\beta$ 
decay, though generally below the reach of planned experiments, could 
approach it in some parameteric regions.
Within the same framework, we also study Yukawa textures with a fewer number of zeros, but with exact 
$\mu\tau$
symmetry. We further formulate the detailed scheme of the explicit breaking of $\mu\tau$ symmetry in terms of 
three small parameters for allowed
four zero textures. The observed sizable mixing between the first and third generations of neutrinos is shown to follow
for a suitable choice of these symmetry breaking parameters.

\begin{center}
 {\bf Invited review, to appear in a special issue of Advances in High Energy Physics (AHEP) on neutrinos}
\end{center}

\tableofcontents
%
\section{Introduction}
\label{sec-07:intro}
The impressive experimental progress from neutrino oscillation studies \cite{07-Tortola:2012te}-\cite{07-Ahn:2012nd} and 
the sharpening \cite{07-Thomas:2009ae,07-Parke:2010} of the cosmological upper
bound on the neutrino mass sum have underscored two fundamental but distinct puzzles. 1) Why are the observed neutrinos
 so ultralight, i.e. with masses in the sub-eV range? 2) Why is the three neutrino mixing pattern of two large and one 
small (but measurable) angles so different from the sequentially small CKM mixing angles of quarks? There is a widespread
feeling that the former is due to some kind of a see-saw mechanism\cite{07-Yanagida:1979as}-\cite{07-Mohapatra:1986bd} yielding ultralight Majorana neutrinos. 
It is our contention that the latter has to do do with zeros in neutrino Yukawa textures plus a broken $\mu\tau$
 symmetry.
Let us start with the simplest scheme of three weakly interacting flavored ultralight neutrinos discarding any possible
light sterile ones mixing with them. We hold that there should be a fundamental principle behind a massless particle, as
with gauge invariance and the photon. Since no such principle is identifiable with any single neutrino, we take each 
to have a nonzero mass. Though there are other types of proposed see-saw mechanisms, such as type-II 
\cite{07-Schechter:1980gr,07-Ma:1998dx},
 type-III \cite{07-Foot:1988aq}, inverse
see-saw \cite{07-Mohapatra:1986bd,07-GonzalezGarcia:1988rw} etc, in a minimalist approach we stick to the original type-I with three heavy right chiral electroweak singlet
neutrinos denoted by the column vector $N_R$.

We next turn to the issue of texture zeros. By a texture we mean a configuration of a Yukawa coupling matrix with some
vanishing elements. Texture zeros have a long history in the quark sector where four zero Yukawa textures 
\cite{07-Fritzsch:2002ga}-\cite{07-Babu:2004tn}
 have had distinguished success in fitting the known quark masses and CKM parameters. The problem is simpler there since 
the Dirac quark mass matrix of a given charge, which is the corresponding Yukawa coupling matrix times the Higgs VEV,
 contains all information about physical quark masses. In the case of see-saw induced ultralight Majorana neutrinos,
the elements of the Dirac mass matrix $M_D$ do not carry all  information about physical neutrino masses. The latter
are contained in the elements of the complex symmetric Majorana neutrino mass matrix $M_\nu$ which is related to $M_D$ 
through the standard see-saw formula. There have been initial as well as continuing efforts 
\cite{07-Xing:2002ta}-\cite{07-Merle:2006du}
 to assume the vanishing of certain elements in $M_\nu$. But, we strongly feel that an occurrence of zeros must be linked
 to some fundamental 
symmetry \cite{07-Grimus:2004hf}-\cite{07-Dev:2011jc} or suppression mechanism \cite{07-Froggatt:1978nt} inherent in the
 Lagrangian itself. It seems more natural then to postulate the 
occurrence of such
zeros in some elements of the neutrino Yukawa coupling matrix (equivalently $M_D$) which appears in the Lagrangian 
\cite{07-Branco:2007nb}-\cite{07-Dighe:2009xj}. There are ways
\cite{07-Hagedorn:2004ba}-\cite{07-Mei:2005qp} to ensure the stability of such zeros under quantum corrections in type-I seesaw
 models.

An important point in the context of texture zeros is that of Weak Basis dependence. Both $M_D$ and $M_\nu$ change 
\cite{07-Branco:2005jr} under general (and different) unitary transformations of the left and right chiral fermion fields. 
In consequence, any Yukawa texture is basis dependent. It is further known that those fermions, which do not couple 
mutually in the Lagrangian, can be simultaneously put into a mass diagonal form by suitable basis transformations. Without loss of generality, we can 
therefore
choose a Weak Basis in which the charged lepton fields $l$ and the very heavy right chiral neutrino fields $N_R$ are
mass diagonal with real masses. The question arises as to how a flavor model, corresponding to a given set of texture zeros
in such a basis, would be recognized in a different basis. It has been shown \cite{07-Branco:2005jr} that the vanishing of certain Weak Basis
invariants would be a hallmark of those zeros. This is also related to the linkage of CP violation at low energies, 
probed in short or long baseline experiments, and at high energies, as relevant to leptogenesis. Though that linkage is a major motivation
for postulating Yukawa texture zeros \cite{07-Branco:2007nb}-\cite{07-Adhikary:2010fa}, it is outside the scope of the present
 review.

In this article we focus on the role of texture zeros, occurring in $M_D$, in understanding the observed pattern of neutrino
masses and mixing angles. More generally, we show how they affect key aspects of low energy neutrino phenomenology. Four
is shown to be the maximum number of such zeros allowed within our framework \cite{07-Branco:2007nb}. We classify all possible four zero textures,
seventy two in total \cite{07-Branco:2007nb}. Then we introduce $\mu\tau$ symmetry \cite{07-Adhikary:2009kz}, 
\cite{Fukuyama:1997ky}-\cite{07-Ghosal:2004qb}
as an invariance under the interchange of flavors $\mu$ ($2$)
and $\tau$ ($3$) in the neutrino sector which is motivated by an automatic prediction of vanishing (maximal) mixing
between the first (second) and third generations of neutrinos. This symmetry reduces the preceding seventy two textures
to four which lead to only two distinct forms of $M_\nu$ whose phenomenological consequences are worked out 
\cite{07-Adhikary:2010fa}-\cite{07-Adhikary:2011pv}. Three 
zero textures with $\mu\tau$ symmetry are also shown to have similar consequences, while textures with a lesser number
of zeros have little predictivity \cite{07-Adhikary:2012}. We then discuss the general explicit 
breaking of $\mu\tau$ symmetry in terms of three
 small parameters
and show, within the lowest order of perturbation in those parameters, that the observed small mixing of first and
 third generations of neutrinos can be explained within our framework \cite{07-Adhikary:2012}. 

In Section \ref{sec-07:frmwk} we set up our formalism. Section \ref{sec-07:mutau} contains the classification of all four zero textures and a 
discussion of 
$\mu\tau$ symmetry. Section \ref{sec-07:mutauph} addresses the consequent phenomenological implications. 
In Section \ref{sec-07:othzero} we discuss the realization 
of other $\mu\tau$ symmetric texture zeros. Section \ref{sec-07:mutaub} contains a general discussion of explicit 
$\mu\tau$ symmetry breaking  and how that fits observation. Finally, in Section \ref{sec-07:conclu} we summarize our conclusions.  
  
\section{Framework and Formalism}
\label{sec-07:frmwk}
The relevant mass terms in our starting Lagrangian are
\begin{eqnarray}
-\mathcal{ L}^m &=& 
\overline{\nu^0_L}~M_D~ N^0_R
+\frac{1}{2}\overline{ {N^0}^C_L}~ M_R ~N^0_R +\overline{l^0_L} ~M_l~ {l}^0_R+h.c.\nonumber\\
&=& \frac{1}{2}\left(\begin{array}{cc}\overline{\nu^0_L} &
    \overline{ {N^0}^C_L}\end{array}\right)\left(\begin{array}{cc} 0 &
      M_D \\ M_D^T & M_R\end{array}\right)\left(\begin{array}{c}{\nu^0}^C_R\\
     N^0_R\end{array}\right)+\overline{l^0_L} ~M_l~ {l}^0_R+h.c.,
\label{eq-07:lagml}
\end{eqnarray}
where we have used the general definition of a conjugate fermion field $\psi^C=\gamma_0C\psi^*$ 
($C$ being the charge conjugation matrix) and the identity 
\begin{eqnarray}
{\bar \psi_L}~m~\psi'_R=\overline{{\psi^\prime}^C_L}~m^T~\psi^C_R.
\label{eq-07:ident}
\end{eqnarray} 
Here $M_R$, $M_D$ and  $M_l$ respectively denote the right chiral complex symmetric Majorana mass 
matrix, the neutrino Dirac mass matrix and the charged lepton mass matrix in a three dimensional family space.
 The superscripts '0' identifies the corresponding fields as flavor eigenstate ones. The complex symmetric 
$6\times 6$ neutrino mass matrix in the second line of (\ref{eq-07:lagml}) is denoted $\mathcal{M}$, i.e.
\begin{eqnarray}
\mathcal{M}=\left(\begin{array}{cc} 0 &
      M_D \\ M_D^T & M_R\end{array}\right).
\label{eq-07:m6}
\end{eqnarray}
The energy scale of $M_R$ is taken to be very high ($> 10^9$ GeV), as compared with the electroweak scale
$v\simeq 246$ GeV.

The complete diagonalization of $\mathcal{M}$ leads to
\begin{eqnarray}
 \mathcal{V}^\dagger \mathcal{M}\mathcal{V}^*={\rm diag}\left(d,~D\right),
\label{eq-07:m6d}
\end{eqnarray}
where $\mathcal{V}$ is a $6\times6$ unitary matrix
\begin{eqnarray}
\mathcal{V}=\left(\begin{array}{cc} K &
      G \\ S & T\end{array}\right)
\label{eq-07:v}
\end{eqnarray}
with $3\times 3$ blocks $K$, $G$, $S$ and $T$. In  (\ref{eq-07:m6d}) $d$ and $D$ are three dimensional 
diagonal mass matrices, each with ultralight and heavy real positive entries respectively:
\begin{eqnarray}
d={\rm diag}\left(m_1,~m_2,~m_3\right),\qquad m_{1,2,3}< 1~{\rm eV},
\label{eq-07:d}
\end{eqnarray}
\begin{eqnarray}
D={\rm diag}\left(M_1,~M_2,~M_3\right),\qquad M_{1,2,3}> 10^{9}~{\rm GeV}.
\label{eq-07:D}
\end{eqnarray}

Charged current interactions can then be written in terms of the semi-weak coupling strength $g$ as well as
the respective ultralight neutrino and heavy neutrino fields $\nu_i$ and $N_i$:
\begin{eqnarray}
\mathcal{L}^{cc} = -\frac{g}{\sqrt 2}\left(\overline{l_{iL}}\gamma_\mu K_{ij}\nu_{jL} + \overline{l_{iL}}\gamma_\mu G_{ij}N_{jL}
\right)W^\mu.
\label{eq-07:lagwk}
\end{eqnarray}
In an excellent approximation, the ultralight neutrino masses and mixing angles can now be obtained from 
\begin{eqnarray}
M_\nu\simeq-M_DM_R^{-1}M_D^T,
\label{eq-07:ss}
\end{eqnarray}
\begin{eqnarray}
U^\dagger M_\nu U^*=d.
\label{eq-07:u}
\end{eqnarray}
Eq. (\ref{eq-07:ss}) is the well-known see-saw formula. We also choose to define the matrix $h_\nu=M_\nu M_\nu^\dagger$ 
and have
\begin{eqnarray}
U^\dagger h_\nu U=d^2.
\label{eq-07:hnud}
\end{eqnarray}

In  (\ref{eq-07:u}), $U$ is the Pontecorvo, Maki, Nakagawa, Sakata (PMNS) matrix admitting the standard parametrization
\begin{equation}
U_{\rm PMNS}= \left(\begin{array}{ccc} c_{12} c_{13}&
                      s_{12} c_{13}&
                      s_{13} e^{-i\delta_D}\\
-s_{12} c_{23}-c_{12} s_{23}s_{13} e^{i\delta_D}&c_{12} c_{23}-
s_{12} s_{23} s_{13} e^{i\delta_D}&
s_{23} c_{13}\\
s_{12} s_{23} -c_{12} c_{23} s_{13} e^{i\delta_D}&
-c_{12} s_{23} -s_{12} c_{23} s_{13} e^{i\delta_D}&
c_{23} c_{13}\end{array}\right)
\left(\begin{array}{ccc}e^{i\alpha_{M_1}/2}&0&0\\
         0&e^{i\alpha_{M_2}/2}&0\\
         0&0&1\end{array}\right)
\label{eq-07:pmns}
\end{equation}
with $c_{ij} = cos\,\theta_{ij}$, $s_{ij} = sin\,\theta_{ij}$, 
and $\delta_D$ ($\alpha_{M_1}$, $\alpha_{M_2}$) being the yet unknown  Dirac (Majorana) phase(s).
 We note for the sake of completeness that
the unitary transformation between the column of mass eigenstate of left chiral neutrino fields $\nu_L$ and the 
corresponding
flavor eigenstate $\nu^0_L$ is
\begin{eqnarray}
\nu^0_L=U\nu_L.
\label{eq-07:fmr}
\end{eqnarray}
The additional approximate relations to keep in mind  are those between $U$ and the submatrices $K$, $G$ of $\mathcal{V}$ and
 $M_D$ of $\mathcal{M}$: 
\begin{eqnarray}
K_{ij}=U_{ij} + O(v^2/{M_R}^2),
\label{eq-07:kur}
\end{eqnarray}
\begin{eqnarray}
M_kG_{jk}=(M_D)_{jk} + O(v^3/{M_R}^2),\quad{k~ \rm not~summed}.
\label{eq-07:gmd}
\end{eqnarray}
Needless to add, we always neglect terms of order $v^2/{M_R}^2$.
 
As mentioned earlier, without loss of generality, we can choose the Weak Basis in which $M_l$ and $M_R$ are
\begin{eqnarray}
M_l={\rm diag}\left(m_e,~m_\mu,~m_\tau\right),
\label{eq-07:mld}
\end{eqnarray}
\begin{eqnarray}
M_R={\rm diag}\left(M_1,~M_2,~M_3\right),
\label{eq-07:mrd}
\end{eqnarray}
with real positive entries. All CP-violating phases, stemming from $\mathcal{M}$, are contained
in the Dirac mass matrix $M_D$ in this Weak Basis. As a consequence of (\ref{eq-07:d}, \ref{eq-07:D}) and (\ref{eq-07:ss}, \ref{eq-07:u}), $M_D$ can
 be written in the Casas-Ibarra
form \cite{07-Casas:2001sr}
\begin{eqnarray}
M_D=iU{\sqrt d}R{\sqrt M_R},
\label{eq-07:cas}
\end{eqnarray}
where $R$ is a complex orthogonal matrix: $R^TR=RR^T=I$. An important comment on $M_D$, following from  (\ref{eq-07:ss}), is
 that our condition of no massless neutrino, i.e. ${\rm det}~M_\nu\ne0$, implies that ${\rm det}~M_D\ne0$. This means
that textures of $M_D$ with one vanishing row or column or with a quartet of zeros (i.e. zeros in $ij,~ lk,~ ik$ 
and $lj$ elements with $i\ne l$ and $k\ne j$ and $l$  $=1,~2$ or $3$) are inadmissible since they make 
${\rm det}~M_D$ vanish. Furthermore, in our Weak Basis, for any nonzero entry in $M_D$ with all other elements in its
row or column being zero, $M_\nu$ from (\ref{eq-07:ss}) develops a block diagonal form that is incompatible with
the observed simultaneous mixing of three neutrinos. The same logic holds for any block diagonal texture of $M_D$. Indeed, 
if any row in a texture of $M_D$ is orthogonal, element by element, to both the others, one neutrino family decouples and
therefore makes such a texture inadmissible. These arguments have been shown \cite{07-Branco:2007nb} to be sufficient to rule out all textures in
 $M_D$ with more than four zeros. Four is then the maximum permitted number of zeros in a neutrino Yukawa texture.
\section{Classification of four zero textures and the role of $\mu\tau$ symmetry}
\label{sec-07:mutau}
In this section we provide the classification of all possible four zero neutrino Yukawa textures 
and forms of the surviving textures, since these details were not given in \cite{07-Branco:2007nb,07-Adhikary:2012kb}.
There are $^9C_4=126$ possible four zero neutrino Yukawa textures which can be classified into four classes \cite{07-Adhikary:2012kb}.
In making this classification, we rule out the orthogonality between any two rows or columns by some artificial 
cancellation; 
orthogonality is to be ensured in terms of a vanishing product, element by element. We can  now enumerate four cases. 

\noindent(i) ${\rm det~M_D}\neq 0$ and one family of neutrinos decouples: $9$ textures.

\noindent
For each texture of $M_D$ here, one row is orthogonal to the other rows. It follows that, in the neutrino mass
matrix $M_\nu$ in our chosen basis with a diagonal $M_R$, one neutrino  family always decouples. So, though all neutrinos
are massive here, these textures are to be discarded. 

\noindent
(ii) ${\rm det~M_D}=0$  and one family of neutrinos decouples: $18$ textures

\noindent 
Here each texture has a vanishing row and there are six such textures for every such row. Such a row  
generates a vanishing mass eigenvalue and the corresponding family decouples. Hence this class is also excluded.  

\noindent
(iii) ${\rm det~M_D}= 0$  and no family decouples: $27$ textures \cite{07-Adhikary:2012kb}

\noindent
Each of $18$  textures in this class has a vanishing column and each of the remaining $9$ has a quartet of zeros, leading
to a vanishing ${\rm det}~M_\nu$. So, this class is rejected.
 
\noindent
(iv) ${\rm det~M_D}\neq 0$  and no family decouples: $72$ textures

\noindent
These remaining textures are allowed by the criteria we have set up.

The retained textures are sub-divided into two categories $A$ and $B$. We wish to elaborate on this categorization
 \cite{07-Branco:2007nb}. 
 Let us consistently use the complex parameters $a_k$, $b_k$ and $c_k$ for elements in $M_D$ belonging to the
$k$th column and the first, second and third rows respectively. The two categories then are as follows. 

\noindent\underline{Category $A$}\\

Here every texture has two mutually orthogonal rows ($i,~j$ say, with $i\ne j$) and the corresponding derived $M_\nu$ has
$(M_\nu)_{ij}=0$. Thus there are $54$ such textures divided into three sub-categories, each containing $18$ textures:
($A1$) those with orthogonal rows $1$ and $2$ which generate $(M_{\nu})_{12}=(M_{\nu})_{21}=0$; 
($A2$) those with orthogonal rows $2$ and $3$ which generate $(M_{\nu})_{23}=(M_{\nu})_{32}=0$; 
($A3$) those with orthogonal rows $1$ and $3$ which generate $(M_{\nu})_{13}=(M_{\nu})_{31}=0$.
The explicit form  of each of the $54$ textures in category $A$ within the three sub-categories is shown in Table \ref{tab-07:tcata}.

\noindent\underline{Category B}\\

There are $18$ textures in this category. Each has two orthogonal columns, while no pair of rows is orthogonal. 
Invariably, then, it turns out that one row ($i$, say) has two zeroes and the other two rows (say $k,~l\ne i$) 
have one zero each. It is now a consequence of (\ref{eq-07:ss}) that, in the derived neutrino mass-matrix
$M_\nu$, we have the relation 
 \begin{eqnarray}
{\rm det~cofactor}~\left[(M_\nu)_{kl}\right]= 0.
  \label{eq-07:catbcof}
 \end{eqnarray}
Once again, one can make three subcategories with six entries each. $B1$ has two zeros
in the first row and one zero in each of the other two rows. $B2$ has two zeros in the second row and one zero in
each of the other two rows. $B_3$ has two zeros in the third row and one zero in each of the other two rows. All $18$ 
textures of Category $B$ are shown in Table \ref{tab-07:tcatb} within the three subcategories.

We now raise the question of $\mu\tau$ symmetry which we had explained in the Introduction.
 This symmetry is evidently invalid for the charged lepton mass terms. However, for elements in the   
 Dirac mass 
matrix $M_D$ of neutrinos, it immediately implies the relations
\begin{eqnarray}
(M_D)_{12}=(M_D)_{13},\qquad(M_D)_{21}=(M_D)_{31},\nonumber\\
(M_D)_{23}=(M_D)_{32},\qquad(M_D)_{22}=(M_D)_{33}.
\label{eq-07:mtmd}
\end{eqnarray}
Moreover, for the masses of the very heavy right-chiral neutrinos, we have
\begin{eqnarray}
(M_R)_{22}=(M_R)_{33},
\label{eq-07:mtmr}
\end{eqnarray}
 a result which is transparent as $M_2=M_3$ in our chosen basis.
\begin{table}[!h]
\begin{tabular}{|c|c|c|}
\hline
\multicolumn{3}{|c|}{{\bf Category $A1$} (Orthogonality between rows $1$ and $2$)}\\
\hline
 $\left(\begin{array}{ccc}0 & a_2 & 0\\ 
             b_1 & 0 & b_3\\
             0 & c_2  & c_3\end{array}\right)$ $\left(\begin{array}{ccc}0 & a_2 & 0\\ 
             b_1 & 0 & b_3\\
             c_1 & c_2  & 0\end{array}\right)$ &  
 $\left(\begin{array}{ccc}0 & 0 & a_3\\ 
             b_1 & b_2 & 0\\
             0 & c_2  & c_3\end{array}\right)$ $\left(\begin{array}{ccc}0 & 0 & a_3\\ 
             b_1 & b_2 & 0\\
             c_1 & 0  & c_3\end{array}\right)$ &
$\left(\begin{array}{ccc} a_1 & 0 & 0\\ 
             0 & b_2 & b_3\\
             c_1 & 0  & c_3\end{array}\right)$ $\left(\begin{array}{ccc} a_1 & 0 & 0\\ 
             0 & b_2 & b_3\\
             c_1 & c_2  & 0\end{array}\right)$ \\
\hline 
$\left(\begin{array}{ccc}0 & a_2 & 0\\ 
             0 & 0 & b_3\\
             c_1 & c_2  & c_3\end{array}\right)$ $\left(\begin{array}{ccc} 0 & a_2 & 0\\ 
             b_1 & 0 & 0\\
             c_1 & c_2  & c_3\end{array}\right)$ &  
 $\left(\begin{array}{ccc} 0 & 0 & a_3\\ 
             b_1 & 0 & 0\\
             c_1 & c_2  & c_3\end{array}\right)$ $\left(\begin{array}{ccc} 0 & 0 & a_3\\ 
             0 & b_2 & 0\\
             c_1 & c_2  & c_3\end{array}\right)$ &
$\left(\begin{array}{ccc} a_1 & 0 & 0\\ 
             0 & 0 & b_3\\
             c_1 & c_2  & c_3\end{array}\right)$ $\left(\begin{array}{ccc} a_1 & 0 & 0\\ 
             0 & b_2 & 0\\
             c_1 & c_2  & c_3\end{array}\right)$ \\
\hline 
$\left(\begin{array}{ccc} a_1 & 0 & a_3\\ 
             0 & b_2 & 0\\
             0 & c_2  & c_3\end{array}\right)$ $\left(\begin{array}{ccc} a_1 & 0 & a_3\\ 
             0 & b_2 & 0\\
             c_1 & c_2  & 0\end{array}\right)$ &  
 $\left(\begin{array}{ccc} a_1 & a_2 & 0\\ 
             0 & 0 & b_3\\
             0 & c_2  & c_3\end{array}\right)$ $\left(\begin{array}{ccc} a_1 & a_2 & 0\\ 
             0 & 0 & b_3\\
             c_1 & 0  & c_3\end{array}\right)$ &
$\left(\begin{array}{ccc} 0 & a_2 & a_3\\ 
             b_1 & 0 & 0\\
             c_1 & 0  & c_3\end{array}\right)$ $\left(\begin{array}{ccc} 0 & a_2 & a_3\\ 
             b_1 & 0 & 0\\
             c_1 & c_2  & 0\end{array}\right)$ \\
\hline 
\multicolumn{3}{|c|}{{\bf Category $A2$} (Orthogonality between rows $2$ and $3$)}\\
\hline
$\left(\begin{array}{ccc} 0 & a_2  & a_3\\
             0 & b_2 & 0\\
             c_1 & 0 & c_3\end{array}\right)$ $\left(\begin{array}{ccc} a_1 & a_2  & 0\\
             0 & b_2 & 0\\
             c_1 & 0 & c_3\end{array}\right)$ &  
 $\left(\begin{array}{ccc} 0 & a_2  & a_3\\ 
             0 & 0 & b_2\\
             c_1 & c_2 & 0\end{array}\right)$ $\left(\begin{array}{ccc} a_1 & 0  & a_3\\ 
             0 & 0 & b_3\\
             c_1 & c_2 & 0\end{array}\right)$ &
$\left(\begin{array}{ccc} a_1 & 0  & a_3\\ 
             b_1 & 0 & 0\\
             0 & c_2 & c_3\end{array}\right)$ $\left(\begin{array}{ccc}a_1 & a_2  & 0\\ 
             b_1 & 0 & 0\\
             0 & c_2 & c_3\end{array}\right)$ \\
\hline
$\left(\begin{array}{ccc} a_1 & a_2  & a_3\\
             0 & b_2 & 0\\
             0 & 0 & c_3\end{array}\right)$ $\left(\begin{array}{ccc} a_1 & a_2  & a_1\\
             0 & b_2 & 0\\
             c_1 & 0 & 0\end{array}\right)$ &  
 $\left(\begin{array}{ccc}a_1 & a_2  & a_3\\ 
             0 & 0 & b_2\\
             0 & c_2 & 0\end{array}\right)$ $\left(\begin{array}{ccc} a_1 & a_2  & a_3\\ 
             0 & 0 & b_3\\
             c_1 & 0 & 0\end{array}\right)$ &
$\left(\begin{array}{ccc} a_1 & a_2  & a_3\\ 
             b_1 & 0 & 0\\
             0 & 0 & c_3\end{array}\right)$ $\left(\begin{array}{ccc} a_1 & a_2  & a_3\\ 
             b_1 & 0 & 0\\
             0 & c_2 & 0\end{array}\right)$ \\
\hline
$\left(\begin{array}{ccc} 0 & a_2  & a_3\\
             b_1 & 0 & b_3\\
             0 & c_2 & 0\end{array}\right)$ $\left(\begin{array}{ccc} a_1 & a_2  & 0\\
             b_1 & 0 & b_3\\
             0 & c_2 & 0\end{array}\right)$ &  
 $\left(\begin{array}{ccc} 0 & a_2  & a_3\\ 
             b_1 & b_2 & 0\\
             0 & 0 & c_3\end{array}\right)$ $\left(\begin{array}{ccc} a_1 & 0  & a_3\\ 
             b_1 & b_2 & 0\\
             0 & 0 & c_3\end{array}\right)$ &
$\left(\begin{array}{ccc}a_1 & 0  & a_3\\ 
             0 & b_2 & b_3\\
             c_1 & 0 & 0\end{array}\right)$ $\left(\begin{array}{ccc}a_1 & a_2  & 0\\ 
             0 & b_2 & b_3\\
             c_1 & 0 & 0\end{array}\right)$ \\
\hline 
\multicolumn{3}{|c|}{{\bf Category $A3$} (Orthogonality between rows $1$ and $3$)}\\
\hline
 $\left(\begin{array}{ccc}0 & a_2 & 0\\ 
             0 & b_2  & b_3\\
             c_1 & 0 & c_3\end{array}\right)$ $\left(\begin{array}{ccc}0 & a_2 & 0\\ 
             b_1 & b_2  & 0\\
             c_1 & 0 & c_3\end{array}\right)$ &  
 $\left(\begin{array}{ccc}0 & 0 & a_3\\ 
             0 & b_2  & b_3\\
             c_1 & c_2 & 0\end{array}\right)$ $\left(\begin{array}{ccc}0 & 0 & a_3\\ 
             b_1 & 0  & b_3\\
             c_1 & c_2 & 0\end{array}\right)$ &
$\left(\begin{array}{ccc}a_1 & 0 & 0\\ 
             b_1 & b_2 & 0\\
             0 & c_2 & c_3\end{array}\right)$ $\left(\begin{array}{ccc}a_1 & 0 & 0\\ 
             b_1 & 0 & b_3\\
             0 & c_2 & c_3\end{array}\right)$ \\
\hline
 $\left(\begin{array}{ccc}0 & a_2 & 0\\ 
             b_1 & b_2  & b_3\\
             0 & 0 & c_3\end{array}\right)$ $\left(\begin{array}{ccc}0 & a_2 & 0\\ 
             b_1 & b_2  & b_3\\
             c_1 & 0 & 0\end{array}\right)$ &  
 $\left(\begin{array}{ccc}0 & 0 & a_3\\ 
             b_1 & b_2  & b_3\\
             0 & c_2 & 0\end{array}\right)$ $\left(\begin{array}{ccc}0 & 0 & a_3\\ 
             b_1 & b_2 & b_3\\
             c_1 & 0 & 0\end{array}\right)$ &
$\left(\begin{array}{ccc}a_1 & 0 & 0\\ 
             b_1 & b_2 & b_3\\
             0 & c_2 & 0\end{array}\right)$ $\left(\begin{array}{ccc}a_1 & 0 & 0\\ 
             b_1 & b_2 & b_3\\
             0 & 0 & c_3\end{array}\right)$ \\
\hline
 $\left(\begin{array}{ccc}a_1 & 0 & a_3\\ 
             0 & b_2  & b_3\\
             0 & c_2 & 0\end{array}\right)$ $\left(\begin{array}{ccc}a_1 & 0 & a_3\\
             b_1 & b_2  & 0\\
             0 & c_2 & 0\end{array}\right)$ &  
 $\left(\begin{array}{ccc}a_1 & a_2 & 0\\ 
             0 & b_2  & b_3\\
             0 & 0 & c_3\end{array}\right)$ $\left(\begin{array}{ccc}a_1 & a_2 & 0\\ 
             b_1 & 0  & b_3\\
             0 & 0 & c_3\end{array}\right)$ &
$\left(\begin{array}{ccc}0 & a_2 & a_3\\ 
             b_1 & b_2 & 0\\
             c_1 & 0 & 0\end{array}\right)$ $\left(\begin{array}{ccc}0 & a_2 & a_3\\ 
             b_1 & 0 & b_3\\
            c_1 & 0 & 0\end{array}\right)$ \\
\hline 
\end{tabular}
\caption{ Four zero Yukawa textures of $M_D$ in Category $A$ with subcategories $A1$, $A2$, and $A3$} 
\label{tab-07:tcata}
\end{table}
\begin{table}[h]
\begin{tabular}{|c|c|c|}
\hline
\multicolumn{3}{|c|}{{\bf Category $B1$}}\\
\hline
\multicolumn{3}{|c|}{{{\bf First row with two zeros, a pair of orthogonal columns, non-orthogonal  rows}}}\\
\hline
 $\left(\begin{array}{ccc}0 & a_2 & 0\\ 
             0 & b_1 & b_3\\
             c_1 & c_2 & 0\end{array}\right)$ $\left(\begin{array}{ccc}0 & a_2 & 0\\ 
             b_1 & b_2 & 0\\
             0 & c_2  & c_3\end{array}\right)$ &  
 $\left(\begin{array}{ccc}0 & 0 & a_3\\ 
             0 & b_2 & b_3\\
             c_1 & 0  & c_3\end{array}\right)$ $\left(\begin{array}{ccc}0 & 0 & a_3\\ 
             b_1 & 0 & b_3\\
             0 & c_2  & c_3\end{array}\right)$ &
$\left(\begin{array}{ccc}a_1 & 0 & 0\\ 
             b_1 & 0 & b_3\\
             c_1 & c_2  & 0\end{array}\right)$ $\left(\begin{array}{ccc}a_1 & 0 & 0\\ 
             b_1 & b_2 & 0\\
             c_1 & 0  & c_3\end{array}\right)$ \\
\hline 
\multicolumn{3}{|c|}{{\bf Category $B2$}}\\
\hline
\multicolumn{3}{|c|}{{\bf  Second row with two zeros, a pair of orthogonal columns, non-orthogonal  rows}}\\
\hline
$\left(\begin{array}{ccc}0 & a_2  & a_3\\
             0 & b_2 & 0\\
             c_1 & c_2 & 0\end{array}\right)$ $\left(\begin{array}{ccc}a_1 & a_2  & 0\\
             0 & b_2 & 0\\
             0 & c_2 & c_3\end{array}\right)$ &  
 $\left(\begin{array}{ccc}0 & a_2  & a_3\\ 
             0 & 0 & b_2\\
             c_1 & 0 & c_3\end{array}\right)$ $\left(\begin{array}{ccc}a_1 & 0  & a_3\\ 
             0 & 0 & b_3\\
             0 & c_2 & c_3\end{array}\right)$ &
$\left(\begin{array}{ccc}a_1 & 0  & a_3\\ 
             b_1 & 0 & 0\\
             c_1 & c_2 & 0\end{array}\right)$ $\left(\begin{array}{ccc}a_1 & a_2  & 0\\ 
             b_1 & 0 & 0\\
             c_1 & 0 & c_3\end{array}\right)$ \\
\hline 
\multicolumn{3}{|c|}{{\bf Category $B3$}}\\
\hline
\multicolumn{3}{|c|}{{\bf Third row with two zeros, a pair of orthogonal columns, non-orthogonal  rows}}\\
\hline
 $\left(\begin{array}{ccc}a_1 & a_2 & 0\\ 
             0 & b_2  & b_3\\
             0 & c_2 & 0\end{array}\right)$ $\left(\begin{array}{ccc} 0 & a_2 & a_3\\
             b_1 & b_2  & 0\\
             0 & c_2 & 0\end{array}\right)$ &  
 $\left(\begin{array}{ccc}a_1 & 0 & a_3\\ 
             0 & b_2  & b_3\\
             0 & 0 & c_3\end{array}\right)$ $\left(\begin{array}{ccc}0 & a_2 & a_3\\ 
             b_1 & 0  & b_3\\
             0 & 0 & c_3\end{array}\right)$ &
$\left(\begin{array}{ccc}a_1 & 0 & a_3\\ 
             b_1 & b_2 & 0\\
             c_1 & 0 & 0\end{array}\right)$ $\left(\begin{array}{ccc} a_1 & a_2 & 0\\ 
             b_1 & 0 & b_3\\
            c_1 & 0 & 0\end{array}\right)$ \\
\hline 
\end{tabular}
\caption{ Four zero Yukawa textures of $M_D$ in Category $B$ with subcategories $B1$, $B2$, and $B3$} 
\label{tab-07:tcatb}
\end{table}
On account of (\ref{eq-07:ss}) and (\ref{eq-07:mtmd}) as well as (\ref{eq-07:mtmr}), one is immediately led to the following relations
among elements of the complex symmetric ultralight neutrino Majorana mass matrix $M_{\nu}$:
\begin{eqnarray}
(M_\nu)_{12}=(M_\nu)_{13}\qquad(M_\nu)_{22}=(M_\nu)_{33}.
\label{eq-07:mtmnu}
\end{eqnarray}
We take these as  statements of a custodial $\mu\tau$ symmetry in the ultralight neutrino sector. One can now invert
(\ref{eq-07:u}) and explore the consequences of (\ref{eq-07:mtmnu}) in the parametrization of (\ref{eq-07:pmns}). An immediate consequence
is the fixing of the two mixing angles pertaining to the third flavor: $\theta_{23}=\pi/4$, $\theta_{13}=0$. Since the
measured former angle is compatible with $45^\circ$ within errors and the latter has been found to be small 
($\simeq 9^\circ$), the occurrence of at least a broken $\mu\tau$ symmetry in nature is a reasonable supposition that we adhere to. 
An interesting footnote to this discussion is the issue of tribimaximal mixing 
\cite{07-Harrison:2002er,07-Harrison:2004uh}
 which subsumes $\mu\tau$ symmetry but
posits the additional relation
\begin{eqnarray}
(M_\nu)_{11}+(M_\nu)_{13}=(M_\nu)_{22}+(M_\nu)_{23},
\label{eq-07:mnutbr}
\end{eqnarray}
leading to a fixation of the remaining mixing angle $\theta_{12}=\sin^{-1}(1/{\sqrt 3})\simeq 35.26^\circ $. However,
we shall not make use of (\ref{eq-07:mnutbr}).

An immediate consequence of the imposition of $\mu\tau$ symmetry, via (\ref{eq-07:mtmd}), is the drastic reduction of the 
seventy two allowed four zero textures of $M_D$ to only four \cite{07-Adhikary:2009kz}. This is seen just by inspection. The allowed $\mu\tau$ 
symmetric textures are the following, each involving only three complex parameters.\\
\underline{Category $A$}
\begin{eqnarray}
 {M_{DA}}_1=\left(\begin{array}{ccc}a_1 & a_2 & a_2\\ 
             0 & 0 & b_1\\
             0 & b_1  & 0\end{array}\right),\qquad {M_{DA}}_2=\left(\begin{array}{ccc}a_1 & a_2 & a_2\\ 
             0 & b_1 & 0\\
             0 & 0  & b_1\end{array}\right),
\label{eq-07:mtmdcata}
\end{eqnarray}
\underline{Category $B$}
\begin{eqnarray}
  {M_{DB}}_1=\left(\begin{array}{ccc}a_1 & 0 & 0\\ 
             b_1 & 0 & b_2\\
             b_1 & b_2  & 0\end{array}\right),\qquad {M_{DB}}_2=\left(\begin{array}{ccc}a_1 & 0 & 0\\ 
             b_1 & b_2 & 0\\
             b_1 & 0  & b_2\end{array}\right).
\label{eq-07:mtmdcatb}
\end{eqnarray}
It may be noted that, in either category, any texture can be obtained from the other by the interchange of rows 
$2$ and $3$ or columns $2$ and $3$. Because of $\mu\tau$ symmetry, this means that  the physical content of the two 
textures in each category is the same. Indeed, by use of (\ref{eq-07:ss}), we obtain the same $M_\nu$ for either. Thus we 
have just two allowed ultralight neutrino Majorana mass matrices
\begin{eqnarray}
  M_{\nu A}= -\left(\begin{array}{ccc}a_1^2/M_1 + 2a_2^2/M_2 & a_2b_1/M_2 & a_2b_1/M_2\cr
                        a_2b_1/M_2 & b_1^2/M_2 &0\cr
                        a_2b_1/M_2 & 0 & b_1^2/M_2\end{array}\right)
\label{eq-07:mnucata}
\end{eqnarray}
and 
\begin{eqnarray}
M_{\nu B} = -\left(\begin{array}{ccc}a_1^2/M_1 & a_1b_1/M_1 & a_1b_1/M_1\cr
                         a_1b_1/M_1 & b_1^2/M_1+b_2^2/M_2 &b_1^2/M_1\cr
                       a_1b_1/M_1 & b_1^2/M_1 & b_1^2/M_1+b_2^2/M_2\end{array}\right) 
 \label{eq-07:mnucatb}
\end{eqnarray}
for categories $A$ and $B$ respectively.
\section{Phenomenology with $\mu\tau$ symmetric four zero Yukawa textures}
\label{sec-07:mutauph}
Given $\mu\tau$ symmetry, one automatically obtains that $\theta_{23}=\pi/4$ and $\theta_{13}=0$. The current $3\sigma$
limits on these are $38.6^\circ<\theta_{23}<53.1^\circ$ and $7.0^\circ<\theta_{13}<10.9^\circ$ \cite{07-Tortola:2012te}. We shall later consider 
a small breaking of $\mu\tau$ symmetry. But, for the moment, let us assume the latter to be the exact. The other mass 
and mixing parameters in the ultralight neutrino sector are kept free. Their experimentally allowed $3\sigma$ ranges will 
be used to constrain the non-zero elements of $M_{\nu A}$ and $M_{\nu B}$ in Table \ref{tab-07:param}. We define 
$\Delta^2_{ij}=m_i^2-m_j^2$ where $i,~j$ ($=1,~2,~3$) refer to the mass eigenstate neutrinos. It will now be 
convenient to reparametrize the elements of $M_{\nu A}$ and  $M_{\nu B}$ in (\ref{eq-07:mnucata}) and
(\ref{eq-07:mnucatb}) respectively in the way given in Table \ref{tab-07:param}. Here $k_1$, $k_2$, $l_1$, $l_2$ are real and positive 
quantities while $\alpha$, $\alpha^\prime$, $\beta$, $\beta^\prime$ are phases. However, the phases $\alpha^\prime$
 and $\beta^\prime$ can be absorbed in the definition of the first family neutrino field $\nu_e$ for $M_{\nu A}$
 and $M_{\nu B}$ respectively and therefore are not physical. Moreover, the overall phase in $m_{A,B}$ can also 
be absorbed by a further redefinition of all flavor eigenstate neutrino fields. So we can treat $m_{A,B}$ as real for further
discussions. In addition, we have defined in Table \ref{tab-07:param} sets of derived real 
quantities $X^{A,B}_{1-4}$ which will be related to various observables.

\begin{table}[ht]
\begin{tabular}{|c|c|}
\hline
{\bf Category $A$} & {\bf Category $B$}\\
\hline
 $M_{\nu A}$ & $M_{\nu B}$\\
\hline
 $
m_A = -b_1^2/M_2,$  & $m_B = -b_2^2/M_2, $\\
$k_1e^{i(\alpha+\alpha^\prime)}=\frac{a_1}{b_1}\sqrt{\frac{M_2}{M_1}},$ & 
$l_1 e^{i\beta^\prime}=\frac{a_1}{b_2}\sqrt{\frac{M_2}{M_1}}$\\
$k_2 e^{i\alpha^\prime}= \frac{a_2}{b_1},$  & $l_2e^{i\beta} = \frac{b_1}{b_2}\sqrt{\frac{M_2}{M_1}}$ \\
$\alpha = {\rm arg}\frac{a_1}{a_2}$ & $\beta = {\rm arg}\frac{b_1}{b_2}$\\
\hline 
$
X_1^A = 2\sqrt{2}k_2{[{(1+2k_2^2)}^2 + k_1^4 + 2k_1^2(1+2k_2^2)\cos2
\alpha]}^{1/2} 
$ & $
X_1^B = 2\sqrt{2}l_1l_2{[{(l_1^2+2l_2^2)}^2 + 
1+ 2(l_1^2+2l_2^2)\cos2\beta]}^{1/2}
$\\
$
X_2^A = 1-k_1^4-4k_2^4-4k_1^2k_2^2\cos2\alpha
$ & $
X_2^B = 1+4l_2^2\cos2\beta+4l_2^4-l_1^4
$\\
$
X_3^A = 1-4k_2^4-k_1^4-4k_1^2k_2^2\cos2\alpha - 4k_2^2
$ & $
X_3^B = 1-{(l_1^2+2l_2^2)}^2 - 4l_2^2\cos2\beta
$\\
$
X_4^A= k_1^4+4k_2^4+4k_1^2k_2^2\cos2\alpha
$ & $
X_4^B = l_1^4
$\\
\hline
\end{tabular}
\caption{ Reparametrized quantities and relevant functions in Category $A$ and Category $B$} 
\label{tab-07:param}
\end{table}  
\begin{table}[]
\begin{center}
\begin{tabular}{|c|c|}
\hline
{\bf Quantity} & {\bf Experimental $3\sigma$ range}\\
\hline
$\Delta_{21}^2$ & $7.12\times 10^{-5}~eV^2<\Delta_{21}^2<8.20\times 10^{-5}~eV^2$\\
$\Delta_{32}^2<0$ & $-2.76\times 10^{-3}~ eV^2<\Delta_{32}^2<-2.22\times 10^{-3}~ eV^2$\\
$\Delta_{32}^2>0$ & $2.18\times 10^{-3}~ eV^2<\Delta_{32}^2<2.70\times 10^{-3}~ eV^2$\\
$\theta_{12}$ & $37.46^\circ<\theta_{12}<31.30^\circ$\\
\hline
\end{tabular}
\end{center}
\caption{Input experimental values \cite{07-Tortola:2012te}} 
\label{tab-07:input}
\end{table}
With the reparametrization given in Table \ref{tab-07:param}, $M_{\nu A}$ and $M_{\nu B}$ assume the simple forms
\begin{eqnarray}
M_{\nu A} = m_A\left(\begin{array}{ccc}k_1^2e^{2i\alpha}+2k_2^2&k_2&k_2\cr
                        k_2 &1& 0\cr
                        k_2&0&1\end{array}\right), \qquad 
M_{\nu B} = m_B \left(\begin{array}{ccc}
             l_1^2&l_1l_2e^{i\beta}&l_1l_2e^{i\beta}\cr
                                    l_1l_2e^{i\beta}&l_2^2e^{2i\beta}+
1&l_2^2e^{2i\beta}\cr
                                    l_1l_2e^{i\beta}&l_2^2e^{2i\beta}
&l_2^2e^{2i\beta}+1
\end{array}\right).
 \label{eq-07:pmnua}
\end{eqnarray}
These lead us, through the diagonalization of the matrix $h_\nu$ of (\ref{eq-07:hnud}), to the relations
\begin{eqnarray}
\Delta_{21}^2 = m^2 X,\quad
\Delta_{32}^2 = \frac{m^2}{2}(X_3 - X),\quad
\tan2\theta_{12} = \frac{X_1}{X_2},
 \label{eq-07:mtres}
\end{eqnarray}
where $m=m_{A,B}$ for category $A,~B$ and 
\begin{eqnarray}
X=\sqrt{X_1^2+X_2^2}.
 \label{eq-07:x}
\end{eqnarray}
One can further make use of (\ref{eq-07:u}) to calculate \cite{07-Adhikary:2011pv} the ultralight masses $m_{1,2,3}$ in terms 
$\Delta_{21}^2$ and also the Majorana phases $\alpha_{M_1}$, $\alpha_{M_2}$, cf. (\ref{eq-07:pmns}), in terms of $m_1,~m_2,~m_3$.
The former are given by
\begin{eqnarray}
m_{1,2} = {\left|\Delta_{21}^2\left(\frac{2-X_3\mp X}{2X}\right)\right|}^{1/2},\qquad
m_3  = \left|\Delta_{21}^2/X\right|^{1/2}
 \label{eq-07:m123}
\end{eqnarray}
and the latter by
\begin{eqnarray}
\cos(\alpha_{M_1}-arg.Z) = 
\frac{|Z|^2m_3^2+m_1^2\sin^4\theta_{12}-
m_2^2\cos^4\theta_{12}}{
2m_1 m_3\sin^2\theta_{12}|Z|},
\nonumber\\
\cos(\alpha_{M_2}-arg.Z) = 
\frac{
|Z|^2m_3^2+m_2^2\cos^4\theta_{12}-
m_1^2\sin^4\theta_{12}}{
2m_2 m_3\cos^2\theta_{12}|Z|}.
 \label{eq-07:majo}
\end{eqnarray}
Here $Z = {[{({\cal M}_\nu)}_{22}+{({\cal M}_\nu)}_{23}]}
{[{({\cal M}_\nu)}_{22}-{({\cal M}_\nu)}_{23}]}^{-1}$. The last quantity of physical interest that we calculate in this 
section is the effective mass $m_{\beta\beta}=|(M_\nu)_{11}|$ appearing in the transition amplitude for the yet unobserved
 neutrinoless nuclear double beta decay. That is given by
\begin{eqnarray}
m_{\beta\beta} = {|\Delta_{21}^2 X_4 X^{-1}|}^{1/2},
 \label{eq-07:dbd}
\end{eqnarray}
with $X_4$ as given in Table \ref{tab-07:param}.

\begin{figure}[!h]
\includegraphics[width=8.0cm,height=8.0cm]{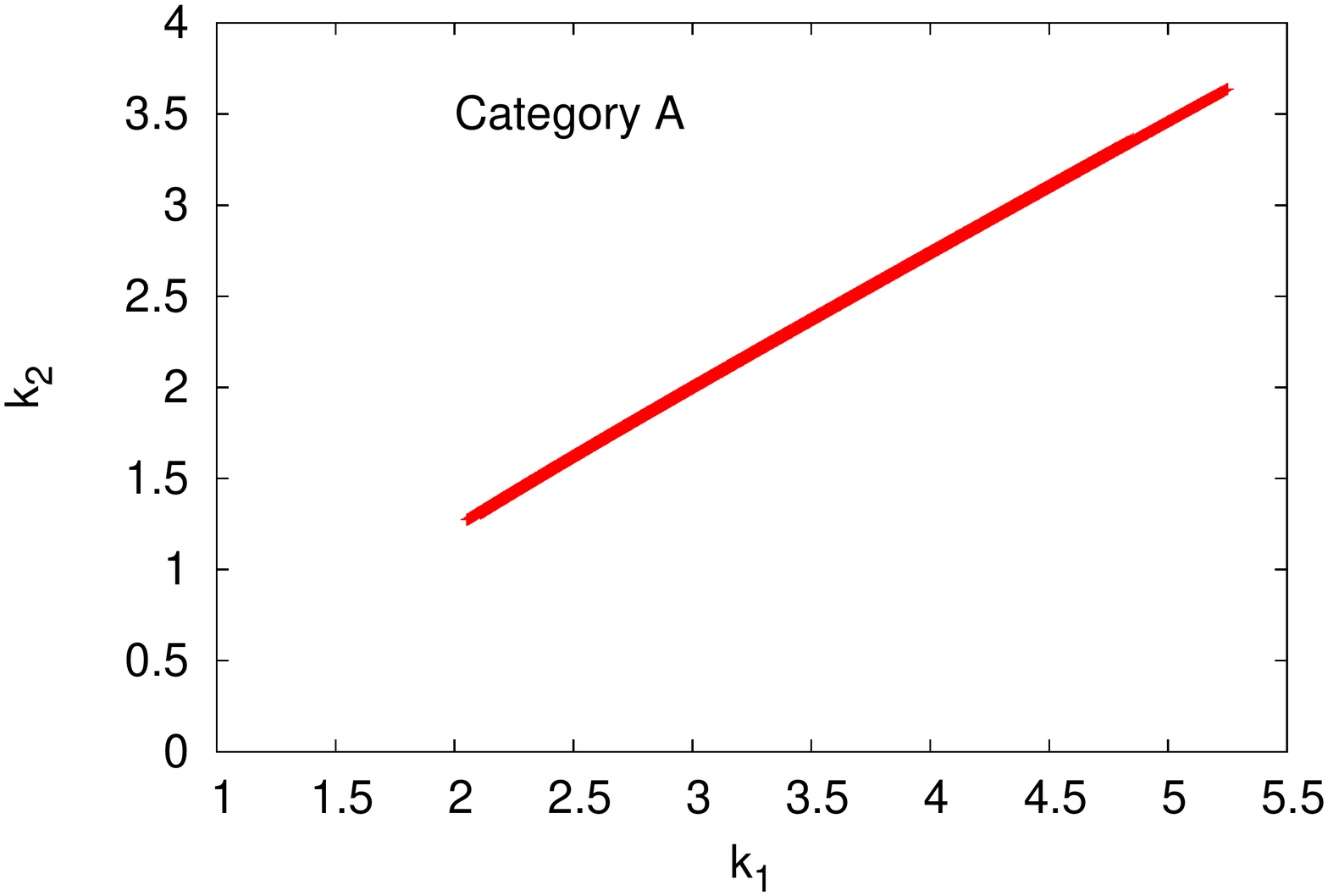}  \includegraphics[width=8.0cm,height=8.0cm]{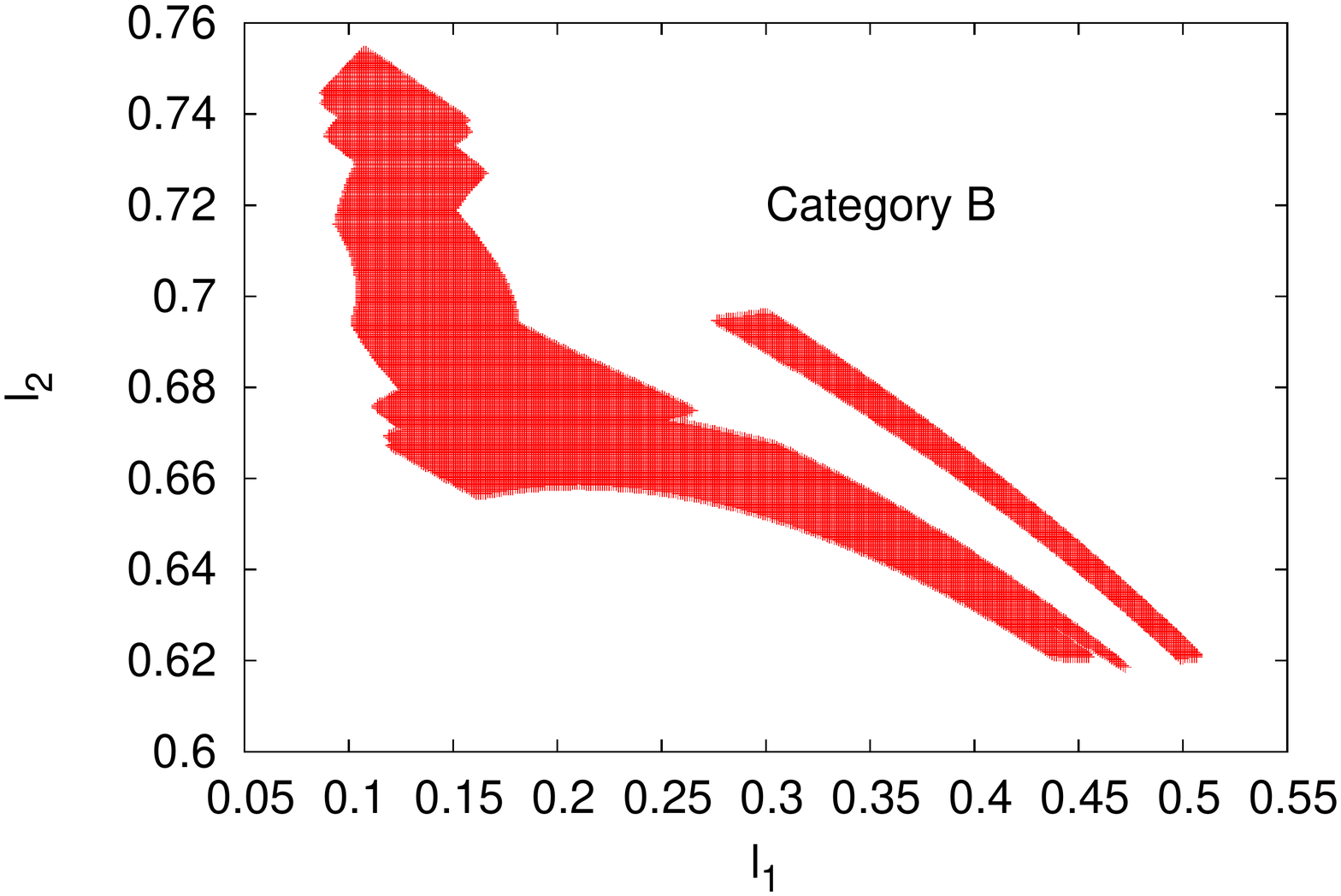}
 \caption{\small
Variation of $k_1$ and $k_2$ in category $A$  and of $l_1$ and $l_2$ in category 
$B$ with $\mu\tau$ symmetry over the $3\sigma$ allowed ranges of 
 $\Delta_{21}^2,~|\Delta_{32}^2|$ and $\theta_{12}$ \cite{07-Adhikary:2009kz}.} 
\label{fig-07:fab}
\end{figure}
Feeding the experimental $3\sigma$ ranges from Table \ref{tab-07:input}, 
 we find that only the inverted mass ordering $\Delta_{32}^2<0$ is allowed in Category
$A$ while only the normal mass ordering $\Delta_{32}^2>0$ is permitted for Category B. Moreover, in the corresponding 
$k_1-k_2/l_1-l_2$ parameter plane \cite{07-Adhikary:2009kz}, very constrained domains are allowed, as shown in 
Fig \ref{fig-07:fab}. 
The phases $\alpha$, $\beta$ are also severely restricted in magnitude, specifically $89.0^\circ\le |\alpha|<90^\circ$ and  
$87.0^\circ\le |\beta|<90^\circ$. These allow just a very limited region in the $X_3-X$ plane, leading to $3\sigma$
lower and upper bounds on the neutrino mass sum $m_1+m_2+m_3$, namely [$0.156,~0.5$] eV/[$0.074,~0.132$] eV for
 an inverted/normal mass-ordering \cite{07-Adhikary:2011pv}. It may be recalled that there is already a lower bound of $0.05$ eV on the said sum from 
atmospheric neutrino data. Furthermore the general consensus \cite{07-Parke:2010} on the least model dependent 
cosmological upper bound
 on it is $0.5$ eV.

\begin{figure}[!h]
\includegraphics[width=5cm,height=5cm]{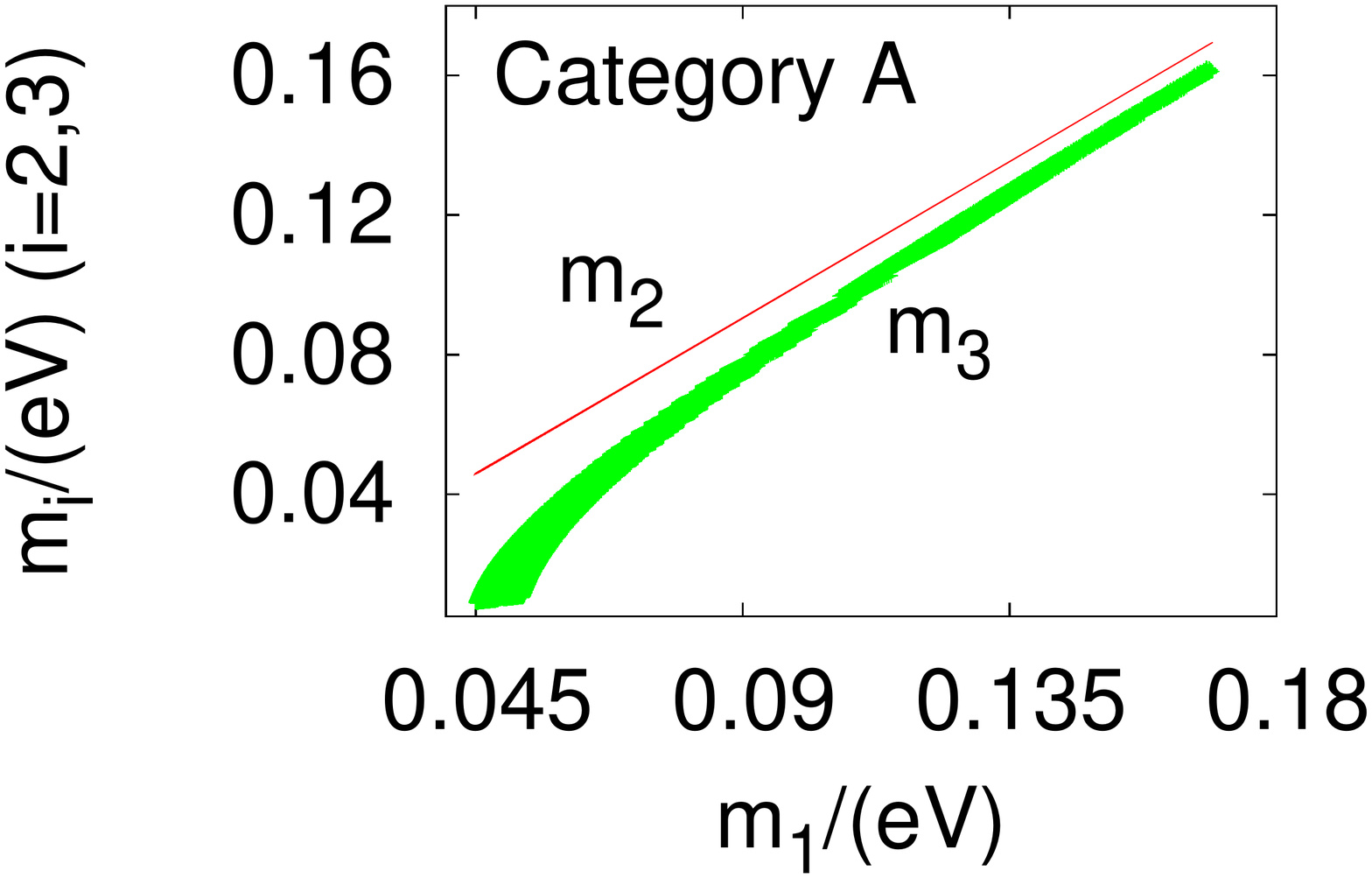}
\includegraphics[width=5cm,height=5cm]{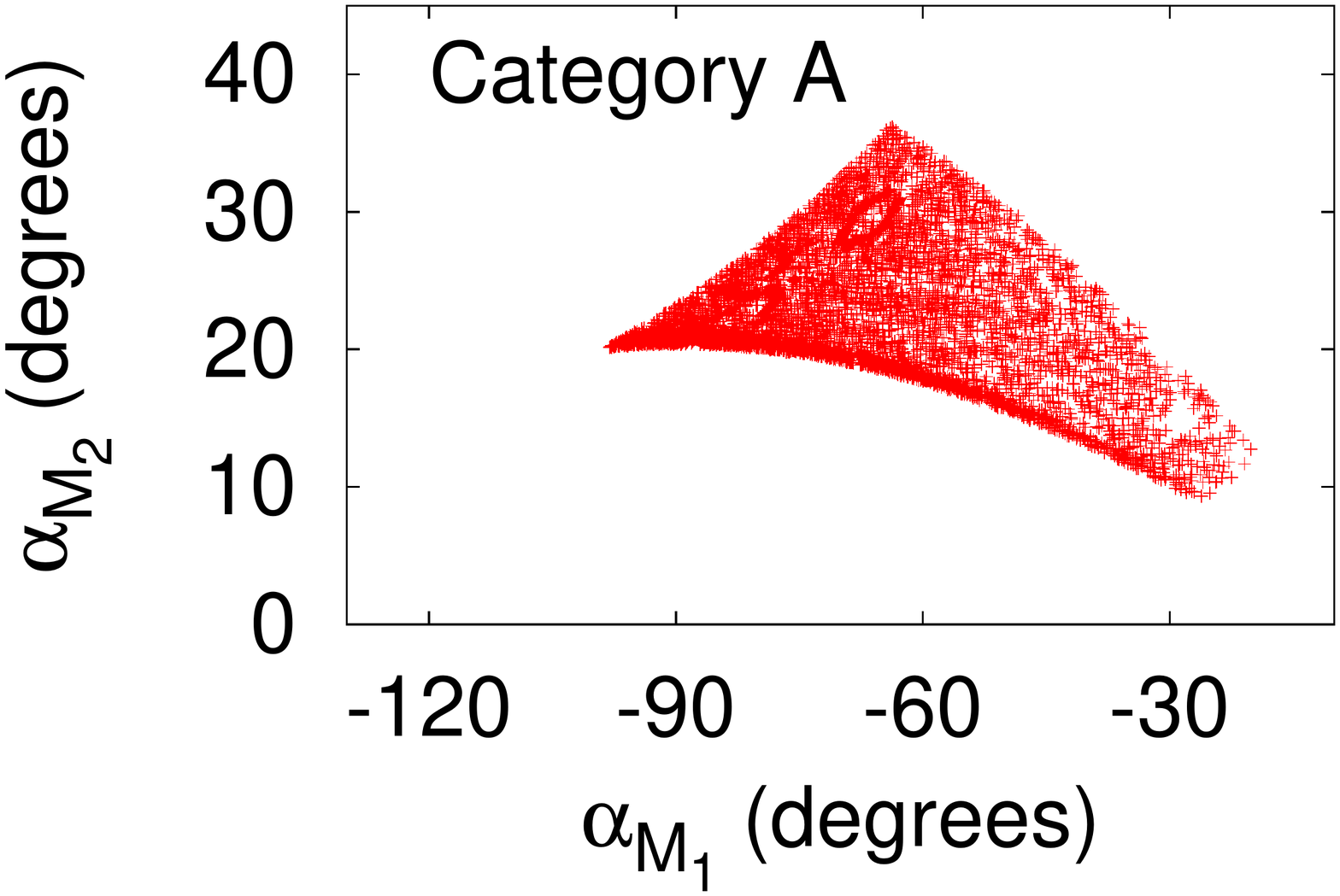}
\includegraphics[width=5cm,height=5cm]{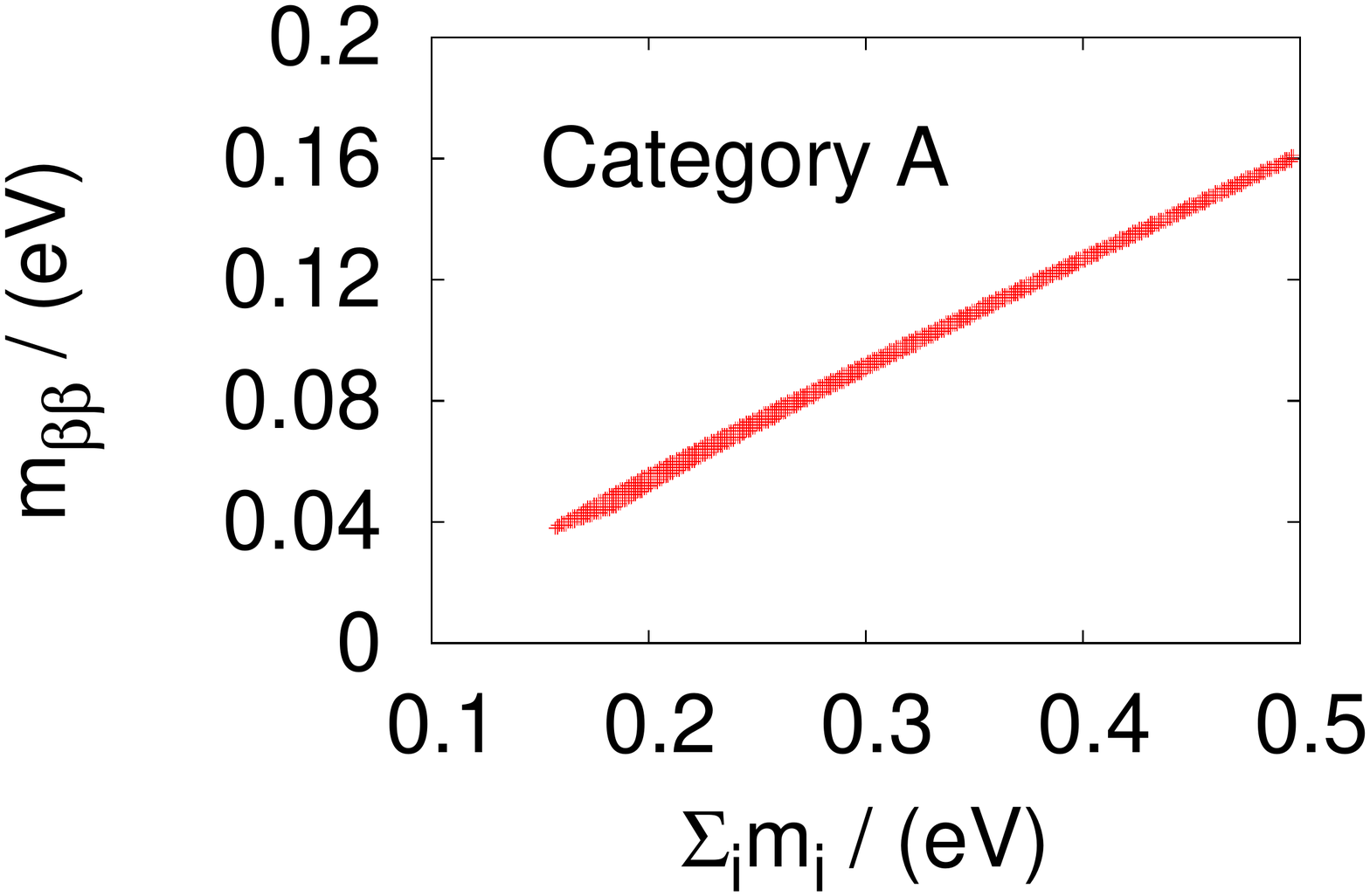}
\vskip 0.1cm
\includegraphics[width=5cm,height=5cm]{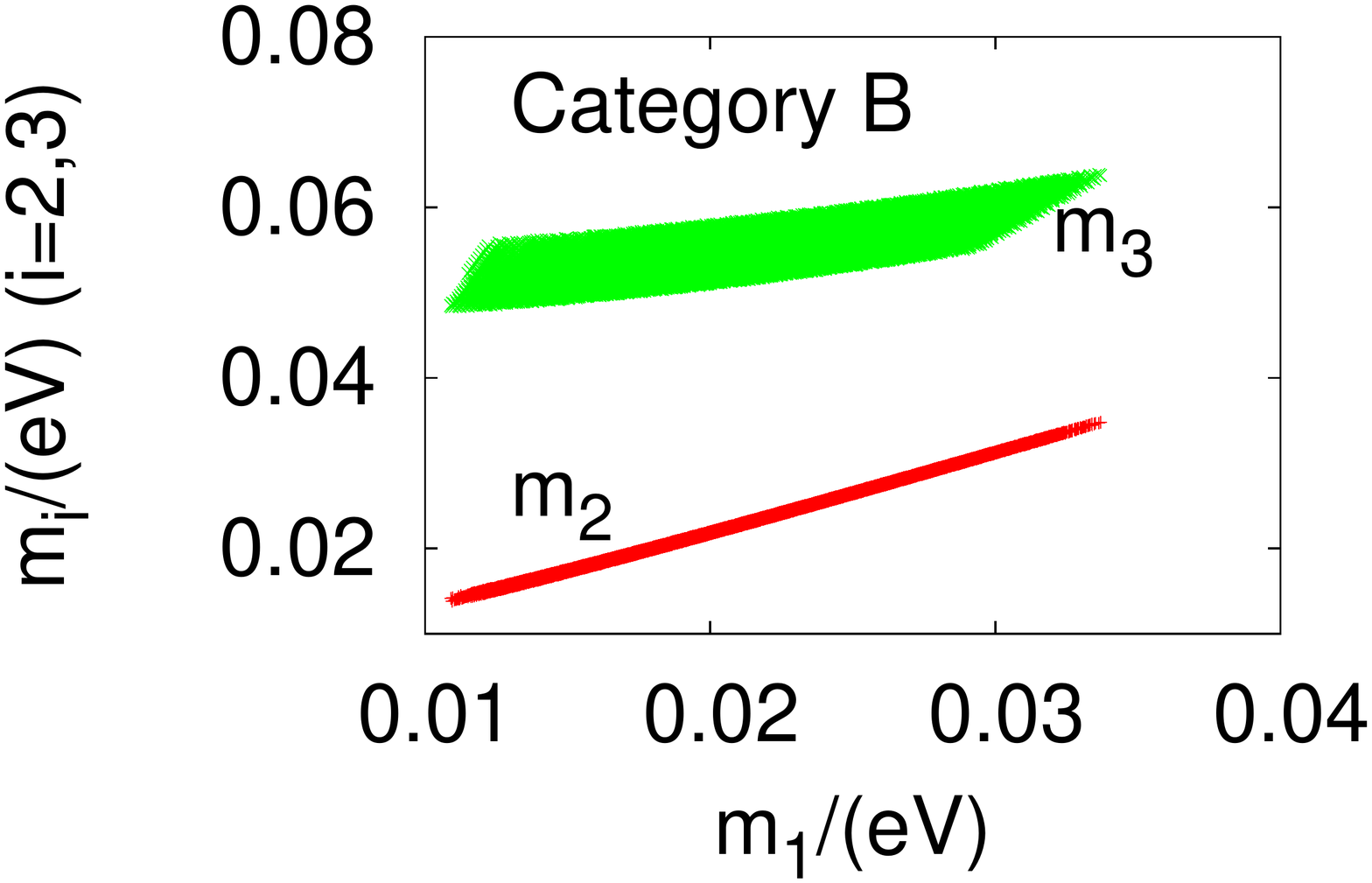}
\includegraphics[width=5cm,height=5cm]{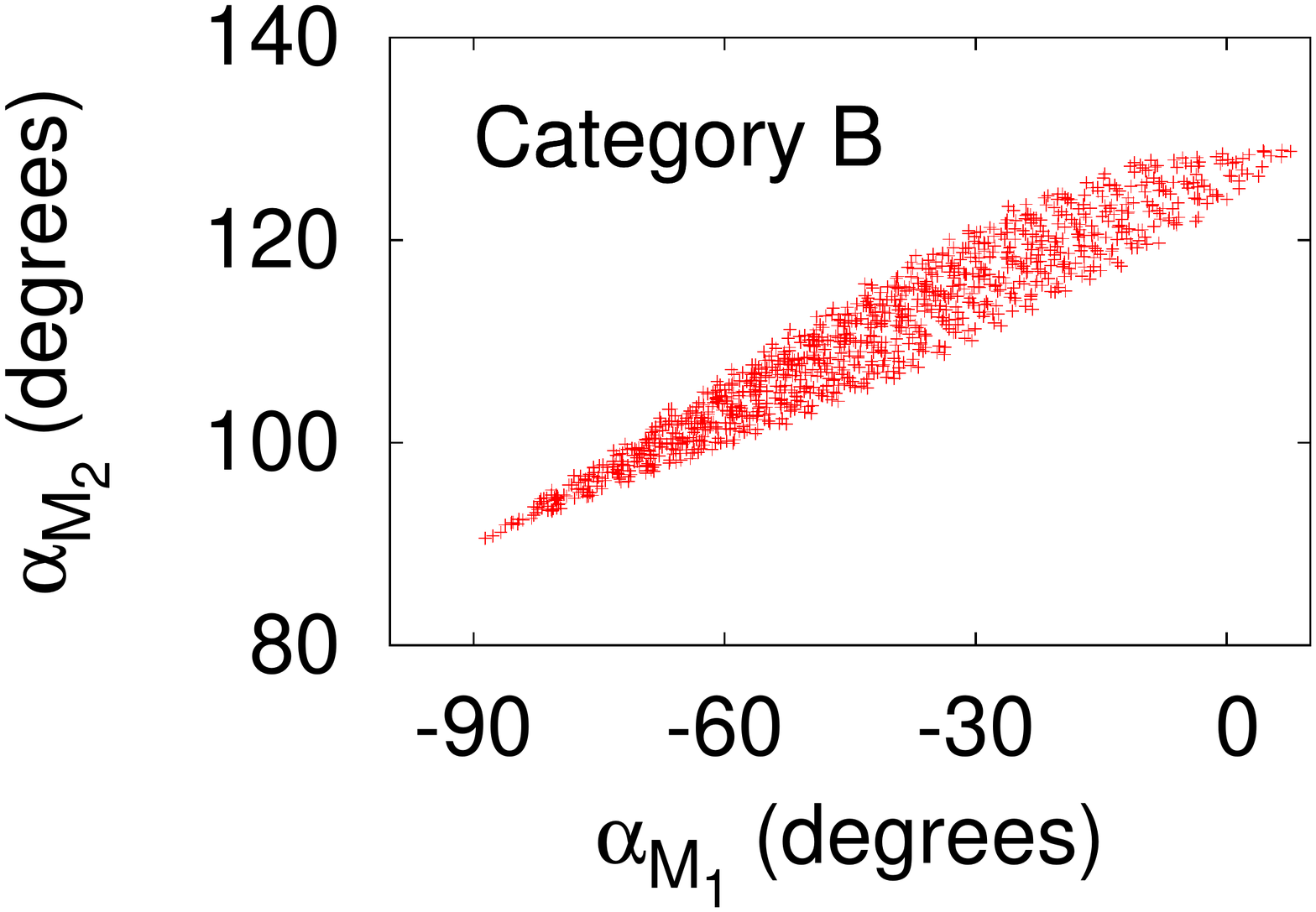}
\includegraphics[width=5cm,height=5cm]{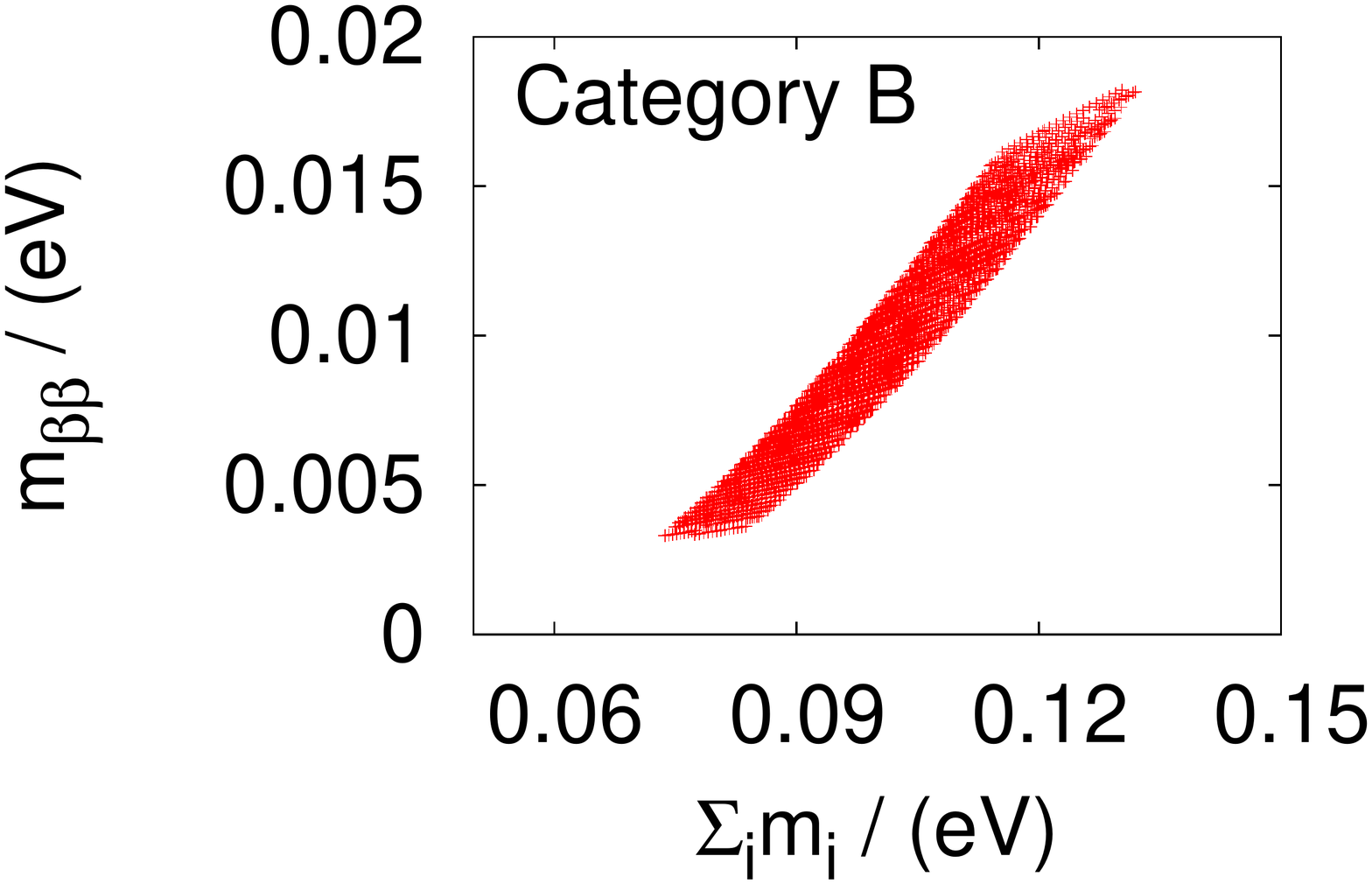}
\caption{\small{{ Allowed $m_{2,3}$ vs $m_1$ (left), 
$\alpha_{M_2}$ vs $\alpha_{M_1}$ (middle) and  
$m_{\beta\beta}$ vs $\Sigma m_i$ (right) for Category $A$ 
(top) and Category $B$ (bottom) \cite{07-Adhikary:2011pv}.}}}
\label{fig-07:pspace}
\end{figure}
Turning to the individual neutrino masses $m_1$/eV, $m_2$/eV and $m_3$/eV respectively, we obtain by use of (\ref{eq-07:m123}) the respective 
$3\sigma$ intervals [$0.0452,~0.1682$], [$0.0457,~0.1684$], [$0.077,~0.1632$] for Category $A$ and  
$[0.0110,~0.0335]$, $[0.0144,~0.0345]$,  
$[0.0485,~0.0638]$ for Category $B$. However, there are correlated constraints among these masses. These are shown in
the left-most panel of Fig \ref{fig-07:pspace}. Given these allowed intervals and correlated constraints, 
it is not possible
 right now to distinguish
between the hierarchical and quasi-degenerate possibilities. But a future reduction of these ranges and domains
 could pin this down. We next come to the Majorana phases $\alpha_{M_1}$, $\alpha_{M_2}$. One can ab initio restrict them
to the interval $-\pi$ to $\pi$ and utilize (\ref{eq-07:majo}) as well as the expressions for $Z$, $m_1$, $m_2$, $m_3$ and
$\tan2\theta_{12}$ in terms of the basic parameters ($k_1$, $k_2$, $\alpha$)/($l_1$, $\l_2$ and $\beta$), depending
on the category. The further application of the phenomenologically acceptable ranges of these parameters, as given above,
leads to the allowed $3\sigma$ intervals $-98.0^\circ\le\alpha_{M_1}\le20.0^\circ$, 
$9.2^\circ\le\alpha_{M_2}\le36.4^\circ$ for Category $A$ and 
$-88.6^\circ\le\alpha_{M_1}\le 7.97^\circ$, $90.7^\circ\le\alpha_{M_2}\le128.8^\circ$ for Category $B$. Allowed values
of $\alpha_{M_1}$ and $\alpha_{M_2}$ are shown in the middle panel of Fig \ref{fig-07:pspace}.

Another quantity to be considered in this section is the double $\beta-$decay effective mass $m_{\beta\beta}$, 
cf. (\ref{eq-07:dbd}). The currently accepted upper bound \cite{07-GomezCadenas:2010gs} on it is $m_{\beta\beta}<0.35$ eV. In comparison, our allowed values
$m_{\beta\beta}$ vs the neutrino mass sum $\sum_i m_i$ are shown in the rightmost panels of of Fig \ref{fig-07:pspace}. 
More absolutely,
we can say that $0.038\le m_{\beta\beta}/{\rm eV}\le 0.161$ for Category $A$ and  
$0.003\le m_{\beta\beta}/{\rm eV}\le 0.0186$ for Category $B$. The region near the upper bound in Category $A$ may be
accessible in forthcoming experiments.

An interesting question pertains to the consequences of the effect of $\mu\tau$ symmetry on couplings 
between the heavy right chiral and the ultralight left chiral neutrinos. The corresponding neutral gauge boson
induced interactions are down by factor $O(v^2/M_R^2)$. On the other hand, the Higgs boson induced interactions affect
leptogenesis modes and have been discussed in detail in \cite{07-Adhikary:2010fa}. Since leptogenesis is outside the
scope of the present article, we do not go into those discussions here.  
\section{Realization of other texture zeros with $\mu\tau$ symmetry}
\label{sec-07:othzero}
Though four is the maximum number of allowed neutrino Yukawa texture zeros, we examine other  textures with 
a lesser number of zeros for completeness \cite{07-Adhikary:2012}. Let us work in the same Weak Basis of real diagonal $M_l$ and $M_R$.
 We wish to study only those textures that are compatible with  $\mu\tau$ symmetry which we believe to be approximately
valid in the real world. The $\mu\tau$ symmetric forms of $M_D$ and $M_R$ now are 
\begin{eqnarray}
M_D=\left(\begin{array}{ccc} a & b & b\\ c & d & e\\c & e & d 
           
          \end{array}
\right),\qquad M_R={\rm diag}~(M_1,~M_2,~M_2)
\label{eq-07:mdmt}
\end{eqnarray}
with $a$, $b$, $c$, $d$, $e$ as complex numbers.\\ 

\noindent\underline{Three zero textures}

\noindent
We first identify possible three zero textures which are compatible with (\ref{eq-07:mdmt}). Apart from $a$, the other 
four complex parameters in $M_D$ come in pairs. So, for any texture with an odd number of zeros, $a$ must vanish. 
For three zero textures the remaining two zeros can be  arranged in $^4C_1=4$ ways. So, the four allowed three
 zero $\mu\tau$ symmetric textures of $M_D$ are:    
\begin{eqnarray}
\left(\begin{array}{ccc} 0 & b & b\\ c & 0 & e\\c & e & 0 \end{array}\right), \qquad 
\left(\begin{array}{ccc} 0 & b & b\\ c & d & 0\\c & 0 & d \end{array}\right),\nonumber\\
\left(\begin{array}{ccc} 0 & 0 & 0\\ c & d & e\\c & e & d \end{array}\right),\qquad
\left(\begin{array}{ccc} 0 & b & b\\ 0 & d & e\\0 & e & d \end{array}\right).
\label{eq-07:md30}
\end{eqnarray}
The last two textures have one vanishing row and one vanishing column respectively. These can be discarded with 
our requirement of no massless neutrino, i.e. ${\rm det}~M_\nu\ne0$, leaving only the first two textures as 
acceptable. In general, there can be $^9C_3=84$ three zero textures. The conditions of (1) $\mu\tau$ symmetry, 
(2) the non-zero value of  ${\rm det}~M_\nu$ and (3) the non-decoupling of any neutrino 
generation reduce this number to only two.  

The first two textures of (\ref{eq-07:md30}) have only three complex parameters each and we can just use $b$, $c$ and $d$ 
for both allowed textures of $M_D$:   
\begin{eqnarray}
\left(\begin{array}{ccc} 0 & b & b\\ c & 0 & d\\c & d & 0 \end{array}\right) \qquad 
\left(\begin{array}{ccc} 0 & b & b\\ c & d & 0\\c & 0 & d \end{array}\right).
\label{eq-07:al30}
\end{eqnarray}
Using the see-saw formula, we obtain an identical form of $M_\nu$ for both textures in  (\ref{eq-07:al30}), namely
\begin{eqnarray}
M_\nu=-\left(\begin{array}{ccc} \frac{2b^2}{M_2} & \frac{bd}{M_2} & \frac{bd}{M_2}\\ 
\frac{bd}{M_2} & \frac{c^2}{M_1}+\frac{d^2}{M_2} & \frac{c^2}{M_1}\\
\frac{bd}{M_2} & \frac{c^2}{M_1} & \frac{c^2}{M_1}+\frac{d^2}{M_2} \end{array}\right).
\label{eq-07:mnu30}
\end{eqnarray}
Eq. (\ref{eq-07:mnu30}) can be written  in the following form under a further reparametrization
\begin{eqnarray}
M_\nu=m_0\left(\begin{array}{ccc} 2k_1^2 e^{i2\alpha_1} &  k_1 e^{i\alpha_1} & k_1 e^{i\alpha_1}\\ 
k_1 e^{i\alpha_1} & k_2^2e^{i\alpha_2}+1 & k_2^2e^{i\alpha_2}\\
k_1 e^{i\alpha_1} & k_2^2e^{i\alpha_2} &  k_2^2e^{i\alpha_2}+1 \end{array}\right),
\label{eq-07:mnu30r}
\end{eqnarray}
with  $m_0=-\frac{d^2}{M_2}$, $k_1e^{i\alpha_1}=\frac{b}{d}$ and 
$k_2e^{i\alpha_2}=\frac{c}{d}{\sqrt \frac{M_2}{M_1}}$. From $M_\nu$, we can remove the phase $\alpha_1$ and 
any phase $\theta_m$
in $m_0$ by rotating $M_\nu$ with the diagonal phase matrix 
${\rm diag}$ ($e^{-i\alpha_1},~1,~1$)$ e^{-i\theta_m/2}$. Thus, $M_\nu$ 
has three real parameters, namely $m_0$, $k_1$ and $k_2$ and only one phase $\alpha_2$. The interesting point to be noted 
is that
the number of independent parameters in $M_\nu$ for $\mu\tau$ symmetric three zero Yukawa textures is the same as that 
for $\mu\tau$ symmetric four zero textures. 
We then have the same phenomenological expressions as in (\ref{eq-07:mtres})-(\ref{eq-07:dbd}), but now with changed definitions 
of $X_{1,2,3,4}$, namely
\begin{eqnarray}
 &&X_1=2{\sqrt 2}k_1\sqrt{(1+2k_1^2)^2+4k_2^4+4k_2^2(1+2k_1^2)\cos2\alpha_2},\nonumber\\
&&X_2=4k_2^4+1+4k_2^2\cos2\alpha_2 -4k_1^4,\nonumber\\
&&X_3=1-4k_1^4-4k_1^2-4k_2^4-4k_2^2\cos2\alpha_2, \nonumber\\
&& X_4= 4k_1^2.
\end{eqnarray}

\noindent\underline{Two zero textures}

Again, looking at the $\mu\tau$ symmetric form of $M_D$ in (\ref{eq-07:mdmt}), we can conclude that, for any even zero textures,
 $a\ne 0$. Two zeros can be fitted to each of the remaining four pairs of parameters in four ways. The four possible two
 zero textures of $M_D$ are:
\begin{eqnarray}
&&\left(\begin{array}{ccc} a & b & b\\ c & 0 & d\\c & d & 0 \end{array}\right), \qquad 
\left(\begin{array}{ccc} a & b & b\\ c & d & 0\\c & 0 & d \end{array}\right),\nonumber\\
&&\left(\begin{array}{ccc} a & 0 & 0\\ c & b & d\\c & d & b \end{array}\right),\qquad
\left(\begin{array}{ccc} a & b & b\\ 0 & c & d\\0 & d & c \end{array}\right)
\label{eq-07:md20}
\end{eqnarray}
and are all allowed. So, the number of allowed $\mu\tau$ symmetric two zero textures is the same as that 
of similar four zero textures.
 We need only the four parameters $a$, $b$, $c$ and $d$ to write down all four textures.
The latter lead to three allowed forms of $M_\nu$; the first
two such textures yield one form  and the remaining two lead to two forms of $M_\nu$.  
These are respectively given by
\begin{eqnarray}
M_\nu=-\left(\begin{array}{ccc} \frac{a^2}{M_1} + \frac{2b^2}{M_2} & \frac{ac}{M_1} + \frac{bd}{M_2} &
 \frac{ac}{M_1} +\frac{bd}{M_2}\\ 
\frac{ac}{M_1} +\frac{bd}{M_2} & \frac{c^2}{M_1}+\frac{d^2}{M_2} & \frac{c^2}{M_1}\\
\frac{ac}{M_1} +\frac{bd}{M_2} & \frac{c^2}{M_1} & \frac{c^2}{M_1}+\frac{d^2}{M_2} \end{array}\right),
\quad 
\label{eq-07:mnu120}
\end{eqnarray}
\begin{eqnarray}
M_\nu=-\left(\begin{array}{ccc} \frac{a^2}{M_1} & \frac{ac}{M_1}  &
 \frac{ac}{M_1} \\ 
\frac{ac}{M_1}  & \frac{c^2}{M_1}+\frac{d^2}{M_2}+\frac{b^2}{M_2} & \frac{c^2}{M_1}+ \frac{2bd}{M_2}\\
\frac{ac}{M_1} & \frac{c^2}{M_1}+ \frac{2bd}{M_2}& \frac{c^2}{M_1}+\frac{d^2}{M_2}+\frac{b^2}{M_2} \end{array}\right),
\quad 
\label{eq-07:mnu220}
\end{eqnarray}
\begin{eqnarray}
M_\nu=-\left(\begin{array}{ccc} \frac{a^2}{M_1} + \frac{2b^2}{M_2} & \frac{bc}{M_2} + \frac{bd}{M_2} &
 \frac{bc}{M_2} +\frac{bd}{M_2}\\ 
\frac{bc}{M_2} +\frac{bd}{M_2} & \frac{c^2}{M_2}+\frac{d^2}{M_2} & \frac{2cd}{M_2}\\
\frac{bc}{M_2} +\frac{bd}{M_2} & \frac{2cd}{M_2} & \frac{c^2}{M_2}+\frac{d^2}{M_2} \end{array}\right).
\quad 
\label{eq-07:mnu320}
\end{eqnarray}
Compared to four and three zero Yukawa textures, there are more independent parameters here. 
Apart from the overall mass scale,
there will be three moduli and two irremovable phases. It is easier to fit the neutrino data with such a larger 
number of parameters
and we do not discuss it any further.

\noindent\underline{One zero}

One zero textures represent the most trivial case among the $\mu\tau$ symmetric neutrino Yukawa texture zeros. 
This is since, as an odd 
zero texture, it must have $a=0$ in (\ref{eq-07:mdmt}). The single allowed texture of $M_D$ is
\begin{eqnarray}
\left(\begin{array}{ccc} 0 & b & b\\ c & d & e\\c & e & d \end{array}\right)
\label{eq-07:md10}
\end{eqnarray}
and yields the following form of $M_\nu$:
\begin{eqnarray}
M_\nu=-\left(\begin{array}{ccc}  \frac{2b^2}{M_2} & \frac{be}{M_2} + \frac{bd}{M_2} &
 \frac{be}{M_2} +\frac{bd}{M_2}\\ 
\frac{be}{M_2} +\frac{bd}{M_2} & \frac{c^2}{M_1}+\frac{d^2}{M_2}+\frac{e^2}{M_2} & \frac{c^2}{M_1}+\frac{2ed}{M_2}\\
\frac{be}{M_2} +\frac{bd}{M_2} & \frac{c^2}{M_1}+\frac{2ed}{M_2} & \frac{c^2}{M_1}+\frac{d^2}{M_2}+\frac{e^2}{M_2} 
\end{array}\right).
\label{eq-07:mnu10}
\end{eqnarray}
Like the two zero textures, this allowed one zero texture has six parameters: one overall real mass scale, three moduli and
two phases. These can easily fit the extant neutrino data.
\section{The breaking of $\mu\tau$ symmetry}
\label{sec-07:mutaub}
As mentioned in  previous sections, the results 
$\t13= 0$ and $\ta23 = \pi/4$ are consequences of the custodial $\mu\tau$ symmetry in $M_\nu$. 
But present neutrino data from
T2K, DOUBLE CHOOZ, RENO and DAYA BAY experiments rule out $\t13= 0$ by $5\sigma$. So, the breaking of $\mu\tau$
 symmetry is an inevitable need
in order to generate a  nonzero value of $\t13$. In addition, a departure of $\ta23 = \pi/4$ would arise from the  
same breaking. Another interesting consequence of a nonzero $\t13$ would be the 
observability of a CKM-type of CP violation in the lepton sector.
Our previous expressions for $\Delta_{21}^2$, $\Delta_{32}^2$ and $\tan2\ts12$ will be modified if $\mu\tau$ symmetry 
is broken.

This symmetry can be broken explicitly or spontaneously or dynamically as with the Renormalization Group evolution 
of Lagrangian parameters. Spontaneous breakdown generally requires the presence of extra scalars
and needs a model with them. We do not go into a discussion of such models here. On the other hand, RG effects 
on neutrino mass can be incorporated
 with the methodology presented in Ref.\cite{07-Dighe:2007ksa,07-Dighe:2006zk}
in terms of the  $\tau$ lepton mass arising through the running of the Yukawa coupling strength from the GUT
scale $\Lambda$ to the weak scale $\lambda$. The effect is characterized by the parameter $\Delta_\tau$ which has the
1-loop expression 
\begin{equation}
\Delta_\tau \simeq \frac{m_\tau^2}{8\pi^2 v^2}{(\tan^2\beta +1)}\ln\left(\frac{\Lambda}{\lambda}
\right), 
\end{equation}
where $\tan\beta$ is the ratio of the VEVs of the up-type and down-type neutral Higgs 
fields in the MSSM and $v^2$ is twice the sum of their squares. Even for a large $\tan\beta$, ( $\tan\beta\simeq 60$)
$\Delta_\tau$ is $O(10^{-3})$ and cannot generate a $\theta_{13}$ of the order of $9^\circ$.

We therefore turn to an explicit breaking of $\mu\tau$ symmetry in $M_D$. That can be realized as the following way:\\
\underline{Category $A$}  
\begin{eqnarray}
{M_{DA}}_1=\left(\begin{array}{ccc}a_1 & a_2 & a_2(1-\epsilon_1e^{i\theta})\\ 
             0 & 0 & b_1(1-\epsilon_2e^{i\phi})\\
             0 & b_1  & 0\end{array}\right),~ {M_{DA}}_2=\left(\begin{array}{ccc}a_1 & a_2 & a_2(1-\epsilon_1e^{i\theta})\\ 
             0 & b_1(1-\epsilon_2e^{i\phi}) & 0\\
             0 & 0  & b_1\end{array}\right),\nonumber\\
\label{eq-07:bmtmdcata}
\end{eqnarray}
\underline{Category $B$}
\begin{eqnarray}
{M_{DB}}_1=\left(\begin{array}{ccc}a_1 & 0 & 0\\ 
             b_1(1-\epsilon_1e^{i\theta}) & 0 & b_2(1-\epsilon_2e^{i\phi})\\
             b_1 & b_2  & 0\end{array}\right),~ {M_{DB}}_2=\left(\begin{array}{ccc}a_1 & 0 & 0\\ 
             b_1(1-\epsilon_1e^{i\theta}) & b_2(1-\epsilon_2e^{i\phi}) & 0\\
             b_1 & 0  & b_2\end{array}\right).\nonumber\\
\label{eq-07:bmtmdcatb}
\end{eqnarray} 
Furthermore, 
\begin{eqnarray}
  M_{R}=\left(\begin{array}{ccc}M_1 & 0 & 0\\ 
             0 & M_2 & 0\\
             0 &  & M_2(1-\epsilon_3)\end{array}\right),
\label{eq-07:bmtmr}
\end{eqnarray} 
where $\epsilon_1e^{i\theta}$, $\epsilon_2e^{i\phi}$ are complex $\mu\tau$ symmetry breaking parameters
($\epsilon_{1,2}$ real) and $\epsilon_3$ is a real $\mu\tau$ symmetry breaking
parameter in $M_R$. For these modified $M_D$  and $M_R$, we have  the following $M_\nu$'s:\\
\underline{Category $A$}
\begin{eqnarray}
&&M_{\nu A1}= m_A\left[\left(\begin{array}{ccc}k_1^2e^{2i\alpha}+2k_2^2&k_2&k_2\cr
                        k_2 &1& 0\cr
                        k_2&0&1\end{array}\right)\right.\nonumber\\
&& -\left.\epsilon_1e^{i\theta}\left(\begin{array}{ccc}2k_2^2&k_2&0\cr
                        k_2 &0& 0\cr
                        0&0&0\end{array}\right)-\epsilon_2e^{i\phi}\left(\begin{array}{ccc}0&k_2&0\cr
                        k_2 &2& 0\cr
                        0&0&0\end{array}\right)-\epsilon_3\left(\begin{array}{ccc}-k_2^2&-k_2&0\cr
                        -k_2 &-1& 0\cr
                        0&0&0\end{array}\right)\right]\nonumber\\
&&M_{\nu A2}= m_A\left[\left(\begin{array}{ccc}k_1^2e^{2i\alpha}+2k_2^2&k_2&k_2\cr
                        k_2 &1& 0\cr
                        k_2&0&1\end{array}\right)\right.\nonumber\\
&& -\left.\epsilon_1e^{i\theta}\left(\begin{array}{ccc}2k_2^2&0&k_2\cr
                        0 &0& 0\cr
                        k_2&0&0\end{array}\right)-\epsilon_2e^{i\phi}\left(\begin{array}{ccc}0&k_2&0\cr
                        k_2 &2& 0\cr
                        0&0&0\end{array}\right)-\epsilon_3\left(\begin{array}{ccc}-k_2^2&0&-k_2\cr
                        0 &0& 0\cr
                        -k_2&0&-1\end{array}\right)\right]\nonumber\\
\label{eq-07:bmnua}
\end{eqnarray}
\underline{Category $B$}
\begin{eqnarray}
M_{\nu B1} &=& m_B \left[\left(\begin{array}{ccc}
             l_1^2&l_1l_2e^{i\beta}&l_1l_2e^{i\beta}\cr
                                    l_1l_2e^{i\beta}&l_2^2e^{2i\beta}+
1&l_2^2e^{2i\beta}\cr
                                    l_1l_2e^{i\beta}&l_2^2e^{2i\beta}
&l_2^2e^{2i\beta}+1
\end{array}\right)\right.\nonumber\\ && \left.-\epsilon_1e^{i\theta}\left(\begin{array}{ccc}
             0&l_1l_2e^{i\beta}&0\cr
                                    l_1l_2e^{i\beta}&2l_2^2e^{2i\beta}&l_2^2e^{2i\beta}\cr
                                    0&l_2^2e^{2i\beta}
&0
\end{array}\right)-\epsilon_2e^{i\phi}\left(\begin{array}{ccc}0&0&0\cr
                        0 &2& 0\cr
                        0&0&0\end{array}\right)-\epsilon_3\left(\begin{array}{ccc}0&0&0\cr
                        0 &-1& 0\cr
                        0&0&0\end{array}\right)\right]\nonumber\\
M_{\nu B2} &=& m_B \left[\left(\begin{array}{ccc}
             l_1^2&l_1l_2e^{i\beta}&l_1l_2e^{i\beta}\cr
                                    l_1l_2e^{i\beta}&l_2^2e^{2i\beta}+
1&l_2^2e^{2i\beta}\cr
                                    l_1l_2e^{i\beta}&l_2^2e^{2i\beta}
&l_2^2e^{2i\beta}+1
\end{array}\right)\right.\nonumber\\ && \left. -\epsilon_1e^{i\theta}\left(\begin{array}{ccc}
             0&l_1l_2e^{i\beta}&0\cr
                                    l_1l_2e^{i\beta}&2l_2^2e^{2i\beta}&l_2^2e^{2i\beta}\cr
                                    0&l_2^2e^{2i\beta}
&0
\end{array}\right)-\epsilon_2e^{i\phi}\left(\begin{array}{ccc}0&0&0\cr
                        0 &2& 0\cr
                        0&0&0\end{array}\right)-\epsilon_3\left(\begin{array}{ccc}0&0&0\cr
                        0 &0& 0\cr
                        0&0&-1\end{array}\right)\right].\nonumber\\
\label{eq-07:bmnub}
\end{eqnarray}
The detailed diagonalization and expressions for mass differences and mixing angles are given in 
Appendix \ref{sec-07:ap1}.

A nonzero $\theta_{13}$  arises after $\mu\tau$ symmetry breaking. The value $\theta_{13}\simeq 9^\circ$ is possible 
for $3\sigma$ variations
of $(\Delta_{21}^2)^{\epsilon_{1,2,3}}$, $(\Delta_{32}^2)^{\epsilon_{1,2,3}}$, $\theta_{12}^{\epsilon_{1,2,3}}$,
$\theta_{23}^{\epsilon_{1,2,3}}$, cf. Appendix. An appropriate choice of symmetry breaking parameters, 
i.e. $\epsilon_{1,2,3}\simeq0.1$, and
 slightly shifted parameter spaces for $k_1,~k_2,~\alpha$ in category $A$
and $l_1,~l_2,~\beta$ in category $B$ are needed. A nonzero $CP$ violating effect can be effected through the Jarlskog invariant 
 $J_{\rm CP}$:
\begin{eqnarray}
J_{\rm CP} = {\rm Im}\frac{H_{12} H_{23} H_{31}}{
(\Delta_{21}^2)^{\epsilon_{1,2,3}} (\Delta_{32}^2)^{\epsilon_{1,2,3}} ({\Delta_{31}^2})^{\epsilon_{1,2,3}}},
 \label{eq-07:CP}
\end{eqnarray}
where elements of $H$ and mass squared differences $(\Delta_{21}^2)^{\epsilon_{1,2,3}}, ~ (\Delta_{32}^2)^{\epsilon_{1,2,3}} $ 
are given in Appendix \ref{sec-07:ap1}. N.B. $({\Delta_{31}^2})^{\epsilon_{1,2,3}}=(\Delta_{21}^2)^{\epsilon_{1,2,3}}+(\Delta_{32}^2)^{\epsilon_{1,2,3}} $.
 A detailed treatment of explicitly broken $\mu\tau$ symmetric four zero and three zero textures is given in 
\cite{07-Adhikary:2012}. 
\section{Concluding summary}
\label{sec-07:conclu}
We have reviewed neutrino Yukawa textures with zeros within the type-I seesaw with three heavy right chiral neutrinos 
and in the basis where 
the latter and the charged leptons are mass diagonal. The conditions of a non-vanishing mass of every ultralight neutrinos
and of the non-decoupling of any neutrino generation allow a maximum of four zeros  
in the neutrino Yukawa coupling matrix $Y_\nu$. There are seventy two such textures. We show that the requirement of an 
exact $\mu\tau$ symmetry, 
coupled with observational constraints,
reduces the {\it seventy two} allowed textures in such a $Y_\nu$ to {\it only four} 
corresponding to {\it just two} different forms of the light neutrino mass matrix $M_{\nu A}/M_{\nu B}$, 
resulting in an inverted/normal mass ordering. Apart from an overall mass scale, $M_\nu$ for every category has
two real parameters and a irremovable phase.
These parameters $k_1,~k_2$ and $|\alpha|$ for Category A and $l_1,~l_2$ and $|\beta|$ for Category B get 
highly restricted, given the $3\sigma$ ranges of measured neutrino mass squared differences and mixing angles.
Neutrino masses and  Majorana phases  
 are predicted within definite ranges with $3\sigma$ laboratory and cosmological inputs. The predicted respective masses 
$m_1$/eV, $m_2$/eV, $m_3$/eV  are [$0.0452,~0.1682$], [$0.0457,~0.1684$], [$0.077,~0.1632$] for Category $A$ and  
$[0.0110,~0.0335]$, $[0.0144,~0.0345]$,  
$[0.0485,~0.0638]$ for Category $B$. Coresponding  intervals of the Majorana phases are 
$-98.0^\circ\le\alpha_{M_1}\le20.0^\circ$, 
$9.2^\circ\le\alpha_{M_2}\le36.4^\circ$ for Category $A$ and 
$-88.6^\circ\le\alpha_{M_1}\le 7.97^\circ$, $90.7^\circ\le\alpha_{M_2}\le128.8^\circ$ for Category $B$.
In addition, we predict the range of the mass scale associated with $0\nu\beta\beta$ 
decay, most of which is well below the reach of planned experiments.
We have also  studied Yukawa textures with a lesser number of zeros, but with exact $\mu\tau$
symmetry. Finally we have formulated the detailed scheme of three parameter explicit breaking of $\mu\tau$ symmetry for allowed
four zero textures. A value of $\theta_{13}\simeq 9^\circ$ can be arranged for a suitable choice of small values
of these symmetry breaking parameters.
\section{Acknowledgments}
\label{sec-07:ack}
The authors would like to thank Prof. Ambar Ghosal for his long collaboration in almost all works discussed in this review.
P.R. acknowledges partial support from a DAE Raja Ramanna Fellowship.
\section{Appendix: Expressions for measurable quantities}
\label{sec-07:ap1}
We can write the general form of a broken $\mu\tau$ symmetric $M_\nu$ in the following way \cite{07-Adhikary:2012}:
\begin{eqnarray}
M_{\nu }= m\left[\left(\begin{array}{ccc}P&Q&Q\cr
                        Q &R& S\cr
                        Q&R&S\end{array}\right) -\epsilon_1e^{i\theta}\left(\begin{array}{ccc}x_1&x_2&x_3\cr
                        x_2 &x_4& x_5\cr
                        x_3&x_5&x_6\end{array}\right)-\epsilon_2e^{i\phi}\left(\begin{array}{ccc}y_1&y_2&y_3\cr
                        y_2 &y_4& y_5\cr
                        y_3&y_5&y_6\end{array}\right)-\epsilon_3\left(\begin{array}{ccc}z_1&z_2&z_3\cr
                        z_2 &z_4& z_5\cr
                        z_3&z_5&z_6\end{array}\right)\right].\nonumber\\
\label{eq-07:bmnug}
\end{eqnarray}
The explicit expressions of $P$, $Q$, $R$, $S$, $x_{1-6}$, $y_{1-6}$ and $z_{1-6}$ for four forms of neutrino mass
matrices after $\mu\tau$ symmetry breaking are given in Table \ref{tab-07:ap}.
\begin{table}[!h]
\begin{tabular}{|c|c|c|c|c|}
\hline
Quanatity&Category $A1$&Category $A2$&Category $B1$ &Category $B2$\\
\hline
$P$&$k_1^2e^{2i\alpha}+2k_2^2$&$k_1^2e^{2i\alpha}+2k_2^2$&$l_1^2$&$l_1^2$\\
\hline 
$Q$&$k_2$&$k_2$&$l_1l_2e^{i\beta}$&$l_1l_2e^{i\beta}$\\
\hline 
$R$&$1$&$1$&$l_2^2e^{2i\beta}+1$&$l_2^2e^{2i\beta}+1$\\
\hline 
$S$&$0$&$0$&$l_2^2e^{2i\beta}$&$l_2^2e^{2i\beta}$\\
\hline 
$x_1$&$2k_2^2$&$2k_2^2$&$0$&$0$\\
\hline 
$x_2$&$k_2$&$0$&$l_1l_2e^{i\beta}$&$l_1l_2e^{i\beta}$\\
\hline 
$x_3$&$0$&$k_2$&$0$&$0$\\
\hline 
$x_4$&$0$&$0$&$2l_2^2e^{2i\beta}$&$2l_2^2e^{2i\beta}$\\
\hline 
$x_5$&$0$&$0$&$l_2^2e^{2i\beta}$&$l_2^2e^{2i\beta}$\\
\hline 
$x_6$&$0$&$0$&$0$&$0$\\
\hline 
$y_1$&$0$&$0$&$0$&$0$\\
\hline 
$y_2$&$k_2$&$k_2$&$0$&$0$\\
\hline 
$y_3$&$0$&$0$&$0$&$0$\\
\hline 
$y_4$&$2$&$2$&$2$&$2$\\
\hline 
$y_5$&$0$&$0$&$0$&$0$\\
\hline 
$y_6$&$0$&$0$&$0$&$0$\\
\hline 
$z_1$&$-k_2^2$&$-k_2^2$&$0$&$0$\\
\hline 
$z_2$&$-k_2$&$0$&$0$&$0$\\
\hline 
$z_3$&$0$&$-k_2$&$0$&$0$\\
\hline 
$z_4$&$-1$&$0$&$-1$&$0$\\
\hline 
$z_5$&$0$&$0$&$0$&$0$\\
\hline 
$z_6$&$0$&$-1$&$0$&$-1$\\
\hline 
\end{tabular}
\caption{Expressions for $P$, $Q$, $R$, $S$, $x_{1-6}$, $y_{1-6}$ and $z_{1-6}$ for categories $A1$, $A2$, $B1$ and $B_4$
from Equations (\ref{eq-07:bmnua}) and (\ref{eq-07:bmnub})} 
\label{tab-07:ap}
\end{table}

 We can now have
\begin{eqnarray}
H = M_\nu M_\nu^\dagger &=& m^2\left[\left(\begin{array}{ccc}
|P|^2 +2|Q|^2& PQ^\star + Q(R^\star+S^\star)&  PQ^\star + Q(R^\star+S^\star)\cr
 P^\star Q + Q^\star(R+S)&|Q|^2+|R|^2+|S|^2&|Q|^2+RS^\star+R^\star S\cr
 P^\star Q + Q^\star(R+S)&|Q|^2+R^*S+RS^*&|Q|^2+|R|^2+|S|^2
\end{array}\right)\right.\nonumber\\
&-&\left.\epsilon_1\left(\begin{array}{ccc}u_1&u_2^*&u_3^*\cr
                        u_2 &u_4& u_5^*\cr
                        u_3&u_5&u_6\end{array}\right)-\epsilon_2\left(\begin{array}{ccc}v_1&v_2^*&v_3^*\cr
                        v_2 &v_4& v_5^*\cr
                        v_3&v_5&v_6\end{array}\right)-\epsilon_3\left(\begin{array}{ccc}w_1&w_2^*&w_3^*\cr
                        w_2 &w_4& w_5^*\cr
                        w_3&w_5&w_6\end{array}\right)\right].
\label{eq-07:bhg}
\end{eqnarray}
Here
\begin{eqnarray}
u_1 &=&\left[P^*x_1+Q^*(x_2+x_3)\right]e^{i\theta}+\left[Px^*_1+Q(x_2^*+x_3^*)\right]e^{-i\theta},\nonumber\\
u_2 &=&\left[P^*x_2+Q^*(x_4+x_5)\right]e^{i\theta}+\left[Qx_1^*+Rx_2^*+Sx_3^*\right]e^{-i\theta},\nonumber\\
u_3 &=&\left[P^*x_3+Q^*(x_5+x_6)\right]e^{i\theta}+\left[Qx_1^*+Sx_2^*+Rx_3^*\right]e^{-i\theta},\nonumber\\
u_4 &=&\left[Q^*x_2+R^*x_4+S^*x_5\right]e^{i\theta}+\left[Qx_2^*+Rx_4^*+Sx_5^*\right]e^{-i\theta},\nonumber\\
u_5 &=&\left[Q^*x_3+R^*x_5+S^*x_6\right]e^{i\theta}+\left[Qx_2^*+Sx_4^*+Rx_5^*\right]e^{-i\theta},\nonumber\\
u_6 &=&\left[Q^*x_3+S^*x_5+R^*x_6\right]e^{i\theta}+\left[Qx_3^*+Sx_5^*+Rx_6^*\right]e^{-i\theta}.
\label{eq-07:6uf}
\end{eqnarray}
\noindent
Note that $v_i$ and $w_i$ have similar functional forms as $u_i$. If we write $u_i$ in the following way
\begin{eqnarray}
u_i =f_i(x_1, x_2,...,x_6, \theta),
\label{eq-07:uf}
\end{eqnarray}
then $v_i$ and $w_i$ will be
\begin{eqnarray}
v_i &=& f_i(y_1, y_2,...,y_6, \phi),\nonumber\\
w_i &=& f_i(z_1, z_2,...,z_6, 0).
\label{eq-07:vwf}
\end{eqnarray}
 
The diagonalization of $H$ yields ${\rm diag.}{(m_1^2, m_2^2, m_3^2)}$ and also expressions for five relevant 
measurable quantities.
 The latter are : $\Delta_{21}^2$, $\Delta_{32}^2$, $\ts12$, $\ta23$ and $\t13$. We will associate superscripts $\epsilon_{1,2,3}$
with all of these five quantities to distinguish them from their unperturbed expressions. The relevant functions for these 
physical quantities
are given below.
\begin{eqnarray}
U_1 &=&\frac{1}{2}\left[-2c_{12}^2u_1+\sqrt{2}c_{12}s_{12}\left\{(u_2+u_3)e^{-i\psi}+(u_2^*+u_3^*)e^{i\psi}\right\}
-s_{12}^2(u_4+u_6+u_5+u_5^*)\right],\nonumber\\
U_2 &=&\frac{1}{2}\left[-\sqrt{2}c_{12}^2(u_2+u_3)e^{-i\psi}+\sqrt{2}s_{12}^2(u_2^*+u_3^*)e^{i\psi}+
c_{12}s_{12}(u_4+u_6-2u_1+u_5+u_5^*)\right],\nonumber\\
U_3 &=&\frac{1}{2}\left[\sqrt{2}c_{12}(u_2-u_3)e^{-i\psi}+s_{12}(u_6-u_4+u_5-u_5^*)\right],\nonumber\\
U_4 &=&\frac{1}{2}\left[-2s_{12}^2u_1-\sqrt{2}c_{12}s_{12}\left\{(u_2+u_3)e^{-i\psi}+(u_2^*+u_3^*)e^{i\psi}\right\}
-c_{12}^2(u_4+u_6+u_5+u_5^*)\right],\nonumber\\
U_5 &=&\frac{1}{2}\left[\sqrt{2}s_{12}(u_2-u_3)e^{-i\psi}-c_{12}(u_6-u_4+u_5-u_5^*)\right],\nonumber\\
U_6 &=&\frac{1}{2}\left[u_5+u_5^*-u_4-u_6\right],
\label{eq-07:6UF}
\end{eqnarray}
where $\psi={\rm arg}~\left[P^*Q+Q^*(R+S)\right]$, $c_{12}=\cos\theta_{12},~s_{12}=\sin\theta_{12}$, 
 $\theta_{12}$ being the unperturbed mixing angle in (\ref{eq-07:mtres}). There are also $V_i$ and $W_i$, $i=1-6$ which
 have similar functional forms
as $U_i$. If we write $U_i$ as
\begin{eqnarray}
U_i =F_i(u_1,~ u_2,...,~u_6),
\label{eq-07:UF}
\end{eqnarray}
then $V_i$ and $W_i$ will be
\begin{eqnarray}
V_i &=& F_i(v_1,~ v_2,...,v_6),\nonumber\\
W_i &=& F_i(w_1,~ w_2,...,~w_6).
\label{eq-07:VWF}
\end{eqnarray}

The final results with three $\mu\tau$ symmetry breaking parameters are
\begin{eqnarray}
(\Delta_{21}^2)^{\epsilon_{1,2,3}} & = & \Delta_{21}^2 + 
m^2\left\{(U_4-U_1)\epsilon_1+(V_4-V_1)\epsilon_2+(W_4-W_1)\epsilon_3
\right\},\nonumber\\
(\Delta_{32}^2)^{\epsilon_{1,2,3}} & = & \Delta_{32}^2 + m^2\left\{(U_6-U_4)\epsilon_1+(V_6-V_4)\epsilon_2+(W_6-W_4)\epsilon_3
\right\},\nonumber\\
(\sin\theta_{12})^{\epsilon_{1,2,3}} & = & \left|s_{12} + c_{12}m^2\left\{\frac{U_2^*\epsilon_1+
V_2^*\epsilon_2+W_2^*\epsilon_3}{\Delta_{21}^2}\right\}\right|,\nonumber\\
(\sin\theta_{23})^{\epsilon_{1,2,3}} & = & \left|\frac{1}{\sqrt 2}
+ \frac{s_{12}m^2}{\sqrt 2}\left\{\frac{U_3^*\epsilon_1+
V_3^*\epsilon_2+W_3^*\epsilon_3}{\Delta_{21}^2+\Delta_{32}^2}\right\}-\frac{c_{12}m^2}{\sqrt 2}\left\{\frac{U_5^*\epsilon_1+
V_5^*\epsilon_2+W_5^*\epsilon_3}{\Delta_{32}^2}\right\}\right|,\nonumber\\
(\sin\theta_{13})^{\epsilon_{1,2,3} }& = &  
\left|c_{12}m^2\left\{\frac{U_3^*\epsilon_1+
V_3^*\epsilon_2+W_3^*\epsilon_3}{\Delta_{21}^2+\Delta_{32}^2}\right\}+s_{12}m^2\left\{\frac{U_5^*\epsilon_1+
V_5^*\epsilon_2+W_5^*\epsilon_3}{\Delta_{32}^2}\right\}\right|.\nonumber\\
\label{eq-07:bmtres}
\end{eqnarray}


\begin{thebibliography}{}
\bibitem{07-Tortola:2012te} 
  D.~V.~Forero, M.~Tortola and J.~W.~F.~Valle,
  ``{\it Global status of neutrino oscillation parameters after recent reactor measurements},''
  arXiv:1205.4018 [hep-ph].

\bibitem{07-Abe:2011fz} 
  Y.~Abe {\it et al.}  [DOUBLE-CHOOZ Collaboration],
  Phys.\ Rev.\ Lett.\  {\bf 108}, 131801 (2012)
  [arXiv:1112.6353 [hep-ex]].

\bibitem{07-Abe:2011sj} 
  K.~Abe {\it et al.}  [T2K Collaboration],
  ``{\it Indication of Electron Neutrino Appearance from an Accelerator-produced Off-axis Muon Neutrino Beam},''
  Phys.\ Rev.\ Lett.\  {\bf 107}, 041801 (2011),
  [arXiv:1106.2822 [hep-ex]].

\bibitem{07-An:2012eh} 
  F.~P.~An {\it et al.}  [DAYA-BAY Collaboration],
  ``{\it Observation of electron-antineutrino disappearance at Daya Bay},''
  Phys.\ Rev.\ Lett.\  {\bf 108}, 171803 (2012)
  [arXiv:1203.1669 [hep-ex]].

\bibitem{07-Ahn:2012nd} 
  J.~K.~Ahn {\it et al.}  [RENO Collaboration],
  ``{\it Observation of Reactor Electron Antineutrino Disappearance in the RENO Experiment},''
  Phys.\ Rev.\ Lett.\  {\bf 108}, 191802 (2012)
  [arXiv:1204.0626 [hep-ex]].

\bibitem{07-Thomas:2009ae} 
  S.~A.~Thomas, F.~B.~Abdalla and O.~Lahav,
  ``{\it Upper Bound of 0.28eV on the Neutrino Masses from the Largest Photometric Redshift Survey},''
  Phys.\ Rev.\ Lett.\  {\bf 105}, 031301 (2010)
  [arXiv:0911.5291 [astro-ph.CO]],
\bibitem{07-Parke:2010} 
 S.~Parke, {\it Unravelling the neutrino mysteries: present 
and future}, Summary talk at NUFACT10, 
http://www.ino.tifr.res.in/nufact2010/proceedings.php/.
\bibitem{07-Yanagida:1979as} 
  T.~Yanagida,
  ``{\it Horizontal Symmetry And Masses Of Neutrinos},''
  Conf.\ Proc.\ C {\bf 7902131}, 95 (1979),
\bibitem{07-GellMann:1980vs} 
  M.~Gell-Mann, P.~Ramond and R.~Slansky,
  ``{\it Complex Spinors And Unified Theories},''
  Conf.\ Proc.\ C {\bf 790927}, 315 (1979),
\bibitem{07-Mohapatra:1979ia} 
  R.~N.~Mohapatra and G.~Senjanovic,
  ``{\it Neutrino Mass and Spontaneous Parity Violation},''
  Phys.\ Rev.\ Lett.\  {\bf 44}, 912 (1980).
\bibitem{07-Schechter:1980gr} 
  J.~Schechter and J.~W.~F.~Valle,
  ``{\it Neutrino Masses in $SU(2) \times U(1)$ Theories},''
  Phys.\ Rev.\ D {\bf 22}, 2227 (1980),
\bibitem{07-Ma:1998dx} 
  E.~Ma and U.~Sarkar,
  ``{\it Neutrino masses and leptogenesis with heavy Higgs triplets},''
  Phys.\ Rev.\ Lett.\  {\bf 80}, 5716 (1998)
  [hep-ph/9802445].
\bibitem{07-Foot:1988aq} 
  R.~Foot, H.~Lew, X.~G.~He and G.~C.~Joshi,
  ``{\it Seesaw Neutrino Masses Induced By A Triplet Of Leptons},''
  Z.\ Phys.\ C {\bf 44}, 441 (1989).
\bibitem{07-Mohapatra:1986bd} 
  R.~N.~Mohapatra and J.~W.~F.~Valle,
  ``{\it Neutrino Mass and Baryon Number Nonconservation in Superstring Models},''
  Phys.\ Rev.\ D {\bf 34}, 1642 (1986),
\bibitem{07-GonzalezGarcia:1988rw} 
  M.~C.~Gonzalez-Garcia and J.~W.~F.~Valle,
  ``{\it Fast Decaying Neutrinos And Observable Flavor Violation In A New Class Of Majoron Models},''
  Phys.\ Lett.\ B {\bf 216}, 360 (1989).
\bibitem{07-Fritzsch:2002ga} 
  H.~Fritzsch and Z.~-z.~Xing,
  ``{\it Four zero texture of Hermitian quark mass matrices and current experimental tests},''
  Phys.\ Lett.\ B {\bf 555}, 63 (2003)
  [hep-ph/0212195].
\bibitem{07-Fritzsch:1999ee} 
  H.~Fritzsch and Z.~-z.~Xing,
  ``{\it Mass and flavor mixing schemes of quarks and leptons},''
  Prog.\ Part.\ Nucl.\ Phys.\  {\bf 45}, 1 (2000)
  [hep-ph/9912358].
\bibitem{07-Babu:2004tn} 
  K.~S.~Babu and J.~Kubo,
  ``{\it Dihedral families of quarks, leptons and Higgses},''
  Phys.\ Rev.\ D {\bf 71}, 056006 (2005)
  [hep-ph/0411226].
\bibitem{07-Xing:2002ta} 
  Z.~-z.~Xing,
  ``{\it Texture zeros and Majorana phases of the neutrino mass matrix},''
  Phys.\ Lett.\ B {\bf 530}, 159 (2002)
\bibitem{07-Frampton:2002yf} 
  P.~H.~Frampton, S.~L.~Glashow and D.~Marfatia,
  ``{\it Zeroes of the neutrino mass matrix},''
  Phys.\ Lett.\ B {\bf 536}, 79 (2002)
  [hep-ph/0201008],
\bibitem{07-Xing:2002ap} 
  Z.~-z.~Xing,
  ``{\it A full determination of the neutrino mass spectrum from two zero textures of the neutrino mass matrix},''
  Phys.\ Lett.\ B {\bf 539}, 85 (2002)
\bibitem{07-Dev:2006qe} 
  S.~Dev, S.~Kumar, S.~Verma and S.~Gupta,
  ``{\it Phenomenology of two-texture zero neutrino mass matrices},''
  Phys.\ Rev.\ D {\bf 76}, 013002 (2007)
  [hep-ph/0612102],
\bibitem{07-Merle:2006du}
  A.~Merle and W.~Rodejohann,
  ``{\it The elements of the neutrino mass matrix: Allowed ranges and  implications
  of texture zeros},''
  Phys.\ Rev.\  D {\bf 73}, 073012 (2006),
  [arXiv:hep-ph/0603111].
\bibitem{07-Grimus:2004hf} 
  W.~Grimus, A.~S.~Joshipura, L.~Lavoura and M.~Tanimoto,
  ``{\it Symmetry realization of texture zeros},''
  Eur.\ Phys.\ J.\ C {\bf 36}, 227 (2004)
  [hep-ph/0405016],
\bibitem{07-Kaneko:2007ea} 
  S.~Kaneko, H.~Sawanaka, T.~Shingai, M.~Tanimoto and K.~Yoshioka,
  ``{\it New Approach to Texture-zeros with S(3) symmetry - Flavor Symmetry and Vacuum Aligned Mass Textures },''
  hep-ph/0703250 [HEP-PH],
\bibitem{07-Dev:2011jc} 
  S.~Dev, S.~Gupta and R.~R.~Gautam,
  ``{\it Zero Textures of the Neutrino Mass Matrix from Cyclic Family Symmetry},''
  Phys.\ Lett.\ B {\bf 701}, 605 (2011)
  [arXiv:1106.3451 [hep-ph]].

\bibitem{07-Froggatt:1978nt} 
  C.~D.~Froggatt and H.~B.~Nielsen,
  ``{\it Hierarchy of Quark Masses, Cabibbo Angles and CP Violation},''
  Nucl.\ Phys.\ B {\bf 147}, 277 (1979).
\bibitem{07-Branco:2007nb} 
  G.~C.~Branco, D.~Emmanuel-Costa, M.~N.~Rebelo and P.~Roy,
  ``{\it Four Zero Neutrino Yukawa Textures in the Minimal Seesaw Framework},''
  Phys.\ Rev.\ D {\bf 77}, 053011 (2008),
  [arXiv:0712.0774 [hep-ph]].
\bibitem{07-Choubey:2008tb}
  S.~Choubey, W.~Rodejohann and P.~Roy,
  ``{\it Phenomenological consequences of four zero neutrino Yukawa textures},''
  Nucl.\ Phys.\  B {\bf 808}, 272 (2009),
  [arXiv:0807.4289 [hep-ph]].
\bibitem{07-Adhikary:2010fa} 
  B.~Adhikary, A.~Ghosal and P.~Roy,
  ``{\it Baryon asymmetry from leptogenesis with four zero neutrino Yukawa textures},''
  JCAP {\bf 1101}, 025 (2011)
  [arXiv:1009.2635 [hep-ph]].
\bibitem{07-Adhikary:2009kz} 
  B.~Adhikary, A.~Ghosal and P.~Roy,
  ``{\it Mu tau symmetry, tribimaximal mixing and four zero neutrino Yukawa textures},''
  JHEP {\bf 0910}, 040 (2009),
  [arXiv:0908.2686 [hep-ph]].
\bibitem{07-Adhikary:2011pv} 
  B.~Adhikary, A.~Ghosal and P.~Roy,
  ``{\it Neutrino Masses, Cosmological Bound and Four Zero Yukawa Textures},''
  Mod.\ Phys.\ Lett.\ A {\bf 26}, 2427 (2011)
  [arXiv:1103.0665 [hep-ph]].
\bibitem{07-Adhikary:2012} 
  B.~Adhikary, A.~Ghosal and P.~Roy,
  ``$\theta_{13}$, $\mu\tau$ symmetry breaking and neutrino Yukawa textures,''
  arXiv:1210.5328 [hep-ph].

\bibitem{Roy:2010jh} 
  P.~Roy,
  ``{\it The Magic of four zero neutrino Yukawa textures},''
  arXiv:1004.0065 [hep-ph].

\bibitem{07-Goswami:2009bd}
  S.~Goswami, S.~Khan and W.~Rodejohann,
  ``{\it Minimal Textures in Seesaw Mass Matrices and their low and high Energy
  Phenomenology},''
  arXiv:0905.2739 [hep-ph].
\bibitem{07-Dighe:2009xj}
  A.~Dighe and N.~Sahu,
  ``{\it Texture zeroes and discrete flavor symmetries in light and heavy Majorana
  neutrino mass matrices: a bottom-up approach},''
  arXiv:0812.0695 [hep-ph].
\bibitem{07-Hagedorn:2004ba} 
  C.~Hagedorn, J.~Kersten and M.~Lindner,
  ``{\it Stability of texture zeros under radiative corrections in see-saw models},''
  Phys.\ Lett.\ B {\bf 597}, 63 (2004)
  [hep-ph/0406103],
\bibitem{07-Antusch:2005gp} 
  S.~Antusch, J.~Kersten, M.~Lindner, M.~Ratz and M.~A.~Schmidt,
  ``{\it Running neutrino mass parameters in see-saw scenarios},''
  JHEP {\bf 0503}, 024 (2005)
  [hep-ph/0501272],
\bibitem{07-Mei:2005qp} 
  J.~-w.~Mei,
  ``{\it Running neutrino masses, leptonic mixing angles and CP-violating phases: From M(Z) to Lambda(GUT)},''
  Phys.\ Rev.\ D {\bf 71}, 073012 (2005),
  [hep-ph/0502015].
\bibitem{07-Branco:2005jr} 
  G.~C.~Branco, M.~N.~Rebelo and J.~I.~Silva-Marcos,
  ``{\it Leptogenesis, Yukawa textures and weak basis invariants},''
  Phys.\ Lett.\ B {\bf 633}, 345 (2006)
  [hep-ph/0510412].

\bibitem{Fukuyama:1997ky} 
  T.~Fukuyama and H.~Nishiura,
  ``{\it Mass matrix of Majorana neutrinos},''
  hep-ph/9702253.


\bibitem{07-GomezIzquierdo:2007vn}
  J. ~C. ~Gomez-Izquierdo and A. ~Perez-Lorenzana,
  ``{\it Softly broken $\mu\leftrightarrow\tau$ symmetry in the minimal see-saw
  model},''
  Phys.\ Rev.\  D {\bf 77}, 113015 (2008),
  [arXiv:0711.0045 [hep-ph]].

\bibitem{07-Baba:2007zzb}
  T.~Baba,
  ``{\it What does mu - tau symmetry imply about leptonic CP violation?},''
  Int.\ J.\ Mod.\ Phys.\  E {\bf 16}, 1373 (2007).

\bibitem{07-Nimai Singh:2007zb}
  N.~Nimai Singh, H.~Zeen Devi and M.~Patgiri,
  ``{\it Phenomenology of neutrino mass matrices obeying $\mu$-$\tau$ reflection
  symmetry},''
  arXiv:0707.2713 [hep-ph].

\bibitem{07-Joshipura:2007sf}
  A.~S.~Joshipura and B.~P.~Kodrani,
  ``{\it Complex CKM matrix, spontaneous CP violation and generalized   $\mu$-$\tau$
  symmetry},''
  Phys.\ Lett.\  B {\bf 670}, 369 (2009),
 [arXiv:0706.0953 [hep-ph]].

\bibitem{07-Adhikary:2006rf}
  B.~Adhikary,
  ``{ \it Soft breaking of L(mu) - L(tau) symmetry: Light neutrino spectrum and
  leptogenesis},''
  Phys.\ Rev.\  D {\bf 74}, 033002 (2006),
  [arXiv:hep-ph/0604009].

\bibitem{07-Baba:2006qb}
  T.~Baba and M.~Yasue,
  ``{\it Correlation between leptonic CP violation and mu - tau symmetry
  breaking},''
  Phys.\ Rev.\  D {\bf 75}, 055001 (2007),
  [arXiv:hep-ph/0612034].

\bibitem{07-Grimus:2006jz}
  W.~Grimus,
  ``{\it Realizations of mu - tau interchange symmetry},''
  arXiv:hep-ph/0610158.

\bibitem{07-Xing:2006xa}
  Z.~z.~Xing, H.~Zhang and S.~Zhou,
  ``{\it Nearly tri-bimaximal neutrino mixing and CP violation from mu - tau
  symmetry breaking},''
  Phys.\ Lett.\  B {\bf 641}, 189 (2006),
  [arXiv:hep-ph/0607091].

\bibitem{07-Haba:2006hc}
  N.~Haba and W.~Rodejohann,
  ``{\it A Supersymmetric Contribution to the Neutrino Mass Matrix and Breaking of
  mu-tau Symmetry},''
  Phys.\ Rev.\  D {\bf 74}, 017701 (2006),
  [arXiv:hep-ph/0603206].

\bibitem{07-Mohapatra:2006un}
  R.~N.~Mohapatra, S.~Nasri and H.~B.~Yu,
  ``{\it Grand unification of mu - tau symmetry},''
  Phys.\ Lett.\  B {\bf 636}, 114 (2006),
  [arXiv:hep-ph/0603020].

\bibitem{07-Ahn:2006nu}
  Y.~H.~Ahn, S.~K.~Kang, C.~S.~Kim and J.~Lee,
  ``{\it Phased breaking of mu - tau symmetry and leptogenesis},''
  Phys.\ Rev.\  D {\bf 73}, 093005 (2006),
 [arXiv:hep-ph/0602160].

\bibitem{07-Fuki:2006ag}
  K.~Fuki and M.~Yasue,
  ``{\it What does mu - tau symmetry imply in neutrino mixings?},''
  Phys.\ Rev.\  D {\bf 73}, 055014 (2006),
 [arXiv:hep-ph/0601118].

\bibitem{07-Nasri:2005gw}
  S.~Nasri,
  ``{\it Implications of mu$\leftrightarrow$tau symmetry on neutrinos and leptogenesis},''
  Int.\ J.\ Mod.\ Phys.\  A {\bf 20}, 6258 (2005).

\bibitem{07-Aizawa:2005yy}
  I.~Aizawa and M.~Yasue,
  ``{\it A new type of complex neutrino mass texture and mu - tau symmetry},''
  Phys.\ Rev.\  D {\bf 73}, 015002 (2006),
  [arXiv:hep-ph/0510132].

\bibitem{07-Mohapatra:2005yu}
  R.~N.~Mohapatra and W.~Rodejohann,
  ``{\it Broken mu-tau Symmetry and Leptonic CP Violation},''
  Phys.\ Rev.\  D {\bf 72}, 053001 (2005),
  [arXiv:hep-ph/0507312].

\bibitem{07-Aizawa:2005ta}
  I.~Aizawa, M.~Ishiguro, M.~Yasue and T.~Kitabayashi,
  ``{\it mu tau permutation symmetry and neutrino mixing for a two-loop radiative
  mechanism},''
  J.\ Korean Phys.\ Soc.\  {\bf 46}, 597 (2005).

\bibitem{07-Kitabayashi:2005fc}
  T.~Kitabayashi and M.~Yasue,
  ``{\it mu - tau symmetry and maximal CP violation},''
  Phys.\ Lett.\  B {\bf 621}, 133 (2005),
 [arXiv:hep-ph/0504212].

\bibitem{07-Mohapatra:2005ra}
  R.~N.~Mohapatra, S.~Nasri and H.~B.~Yu,
  ``{\it Leptogenesis, $\mu-\tau$ Symmetry and $\theta_{13}$},''
  Phys.\ Lett.\  B {\bf 615}, 231 (2005),
  [arXiv:hep-ph/0502026].

\bibitem{07-Mohapatra:2004hta}
  R.~N.~Mohapatra and S.~Nasri,
  ``{\it Leptogenesis and mu - tau symmetry},''
  Phys.\ Rev.\  D {\bf 71}, 033001 (2005),
  [arXiv:hep-ph/0410369].

\bibitem{07-Mohapatra:2004mf}
  R.~N.~Mohapatra,
  ``{\it $\theta_{13}$ as a probe of mu $\leftrightarrow$ tau symmetry for leptons},''
  JHEP {\bf 0410}, 027 (2004),
  [arXiv:hep-ph/0408187].

\bibitem{07-Aizawa:2004qf}
  I.~Aizawa, M.~Ishiguro, T.~Kitabayashi and M.~Yasue,
  ``{\it Bilarge neutrino mixing and mu - tau permutation symmetry for two-loop
  radiative mechanism},''
  Phys.\ Rev.\  D {\bf 70}, 015011 (2004),
  [arXiv:hep-ph/0405201].
\bibitem{07-deMedeirosVarzielas:2005ax}
  I.~de Medeiros Varzielas and G.~G.~Ross,
  ``{\it $SU(3)$ family symmetry and neutrino bi-tri-maximal mixing},''
  Nucl.\ Phys.\  B {\bf 733}, 31 (2006),
  [arXiv:hep-ph/0507176].

\bibitem{07-Harrison:2002et}
  P.~F.~Harrison and W.~G.~Scott,
  ``{\it mu - tau reflection symmetry in lepton mixing and neutrino oscillations},''
  Phys.\ Lett.\  B {\bf 547}, 219 (2002),
  [arXiv:hep-ph/0210197].

\bibitem{07-Kitabayashi:2002jd}
  T.~Kitabayashi and M.~Yasue,
  ``{\it $S(2L)$ permutation symmetry for left-handed mu and tau families and neutrino
  oscillations in an SU(3)L x U(1)N gauge model},''
  Phys.\ Rev.\  D {\bf 67}, 015006 (2003),
 [arXiv:hep-ph/0209294].

\bibitem{07-Ghosal:2004qb}
  A.~Ghosal,
  ``{\it A neutrino mass model with reflection symmetry},''
  Mod.\ Phys.\ Lett.\  A {\bf 19}, 2579 (2004).

\bibitem{07-Casas:2001sr} 
  J.~A.~Casas and A.~Ibarra,
  ``{\it Oscillating neutrinos and muon$\rightarrow$e, gamma},''
  Nucl.\ Phys.\ B {\bf 618}, 171 (2001),
  [hep-ph/0103065].

\bibitem{07-Adhikary:2012kb} 
  B.~Adhikary, M.~Chakraborty and A.~Ghosal,
  ``{\it Scaling ansatz, four zero Yukawa textures and large $\theta_{13}$},''
 Phys.\ Rev.\ D {\bf 86}, 013015 (2012), arXiv:1205.1355 [hep-ph].

\bibitem{07-Harrison:2002er}
  P.~F.~Harrison, D.~H.~Perkins and W.~G.~Scott,
  ``{\it Tri-bimaximal mixing and the neutrino oscillation data},''
  Phys.\ Lett.\  B {\bf 530}, 167 (2002),
 [arXiv:hep-ph/0202074].

\bibitem{07-Harrison:2004uh}
  P.~F.~Harrison and W.~G.~Scott,
  ``{\it Status of tri- / bi-maximal neutrino mixing},''
  arXiv:hep-ph/0402006.
\bibitem{07-GomezCadenas:2010gs} 
  J.~J.~Gomez-Cadenas, J.~Martin-Albo, M.~Sorel, P.~Ferrario, F.~Monrabal, J.~Munoz-Vidal, P.~Novella and A.~Poves,
  ``{\it Sense and sensitivity of double beta decay experiments},''
  JCAP {\bf 1106}, 007 (2011)
  [arXiv:1010.5112 [hep-ex]].

\bibitem{07-Dighe:2007ksa} 
  A.~Dighe, S.~Goswami and P.~Roy,
  ``{\it Radiatively broken symmetries of nonhierarchical neutrinos},''
  Phys.\ Rev.\ D {\bf 76}, 096005 (2007)
  [arXiv:0704.3735 [hep-ph]].
\bibitem{07-Dighe:2006zk} 
  A.~Dighe, S.~Goswami and P.~Roy,
  ``{\it Quark-lepton complementarity with quasidegenerate Majorana neutrinos},''
  Phys.\ Rev.\ D {\bf 73}, 071301 (2006)
  [hep-ph/0602062].
\end{thebibliography}
 \end{document}